\documentclass[twocolumn]{aastex63}

\usepackage{listings}
\usepackage{xcolor}

\definecolor{codegreen}{rgb}{0,0.6,0}
\definecolor{codeblue}{rgb}{0,0,1}
\definecolor{codegray}{rgb}{0.5,0.5,0.5}
\definecolor{codepurple}{rgb}{0.58,0,0.82}
\definecolor{backcolour}{rgb}{0.97,0.97,0.97}

\lstdefinestyle{mystyle}{
  backgroundcolor=\color{backcolour},   commentstyle=\color{codegreen},
  keywordstyle=\color{magenta},
  numberstyle=\tiny\color{codegray},
  stringstyle=\color{codepurple},
  basicstyle=\ttfamily\footnotesize,
  breakatwhitespace=false,         
  breaklines=true,                 
  captionpos=b,                    
  keepspaces=true,                 
  numbers=left,                    
  numbersep=5pt,                  
  showspaces=false,                
  showstringspaces=false,
  showtabs=false,                  
  tabsize=2
}

\newcommand{\A}{\text{\normalfont\AA}}

\def \ebmv{E(B-V)}

\def \lya{Ly$\alpha$}

\def \h2{{\rm H_{2}}}

\def \oiii{[\ion{O}{3}]}

\def \dn4000{D_{{\rm n}}(4000) }

\received{}
\revised{}
\accepted{}
\submitjournal{ApJ}

\shorttitle{Joint Survey Processing}
\shortauthors{JSP Team}
\graphicspath{{./}{figures/}}
\begin{document}

%\title{Joint Survey Processing of {\it Hubble} and {\it Subaru}: Quasars at $z\sim6.5$ and M-dwarfs}
\title{\sc \large  Joint Survey Processing I: Compact oddballs\\ in the COSMOS field $-$ low-luminosity Quasars at $z>6$?}
\shorttitle{Joint Survey Processing I: Compact oddballs in the COSMOS field}

\correspondingauthor{Andreas L. Faisst}
\email{afaisst@ipac.caltech.edu}

\author[0000-0002-9382-9832]{Andreas L. Faisst}
\affiliation{Caltech/IPAC, MS314-6, 1200 E. California Blvd. Pasadena, CA 91125, USA}

\author[0000-0001-7583-0621]{Ranga Ram Chary}
\affiliation{Caltech/IPAC, MS314-6, 1200 E. California Blvd. Pasadena, CA 91125, USA}

\author[0000-0001-9309-0102]{Sergio Fajardo-Acosta}
\affiliation{Caltech/IPAC, MS314-6, 1200 E. California Blvd. Pasadena, CA 91125, USA}

\author[0000-0002-5158-243X]{Roberta Paladini}
\affiliation{Caltech/IPAC, MS314-6, 1200 E. California Blvd. Pasadena, CA 91125, USA}

\author[0000-0001-7648-4142]{Benjamin Rusholme}
\affiliation{Caltech/IPAC, MS314-6, 1200 E. California Blvd. Pasadena, CA 91125, USA}

\author[0000-0003-0987-5738]{Nathaniel Stickley}
\affiliation{Caltech/IPAC, MS314-6, 1200 E. California Blvd. Pasadena, CA 91125, USA}

\author[0000-0003-3367-3415]{George Helou}
\affiliation{Caltech/IPAC, MS314-6, 1200 E. California Blvd. Pasadena, CA 91125, USA}

\author[0000-0003-1614-196X]{John R. Weaver}
\affil{Cosmic Dawn Center (DAWN), Copenhagen, Denmark}
\affil{Niels Bohr Institute, University of Copenhagen, Jagtvej 128, 2200 Copenhagen, Denmark}

\author[0000-0003-2680-005X]{Gabriel Brammer}
\affil{Cosmic Dawn Center (DAWN), Copenhagen, Denmark}
\affil{Niels Bohr Institute, University of Copenhagen, Jagtvej 128, 2200 Copenhagen, Denmark}

\author[0000-0002-6610-2048]{Anton M. Koekemoer}
\affiliation{Space Telescope Science Institute, 3700 San Martin Dr., Baltimore, MD 21218, USA}

\author[0000-0001-7964-9766]{Hironao Miyatake}
\affiliation{Institute for Advanced Research, Nagoya University, Nagoya 464-8601, Japan}
\affiliation{Division of Particle and Astrophysical Science, Graduate School of Science, Nagoya University, Nagoya 464-8602, Japan}
\affiliation{Kavli Institute for the Physics and Mathematics of the Universe (Kavli IPMU, WPI), University of Tokyo, Chiba 277-8582, Japan}

%% Note that the \and command from previous versions of AASTeX is now
%% depreciated in this version as it is no longer necessary. AASTeX 
%% automatically takes care of all commas and "and"s between authors names.

%% AASTeX 6.3 has the new \collaboration and \nocollaboration commands to
%% provide the collaboration status of a group of authors. These commands 
%% can be used either before or after the list of corresponding authors. The
%% argument for \collaboration is the collaboration identifier. Authors are
%% encouraged to surround collaboration identifiers with ()s. The 
%% \nocollaboration command takes no argument and exists to indicate that
%% the nearby authors are not part of surrounding collaborations.

%% Mark off the abstract in the ``abstract'' environment. 
\begin{abstract}

The faint-end slope of the quasar luminosity function at $z\sim6$ and its implication on the role of quasars in reionizing the intergalactic medium at early times has been an outstanding problem for some time. The identification of faint high-redshift quasars with luminosities of $<10^{44.5}$ erg\,s$^{-1}$ is challenging. They are rare (few per square degree) and the separation of these unresolved quasars from late-type stars and compact star-forming galaxies is difficult from ground-based observations alone. In addition, source confusion becomes significant at $>25\,{\rm mag}$, with $\sim30\%$ of sources having their flux contaminated by foreground objects when the seeing resolution is $\sim$0.7$\arcsec$.
We mitigate these issues by performing a pixel-level joint processing of ground and space-based data from Subaru/{\it HSC} and \textit{HST}/ACS. We create a deconfused catalog over the $1.64\,{\rm deg^2}$ of the COSMOS field, after accounting for spatial varying PSFs and astrometric differences between the two datasets.
We identify twelve low-luminosity (M$_{UV}\sim-21\,{\rm mag}$) $z>6$ quasar candidates through \textit{(i)} their red color measured between ACS/F814W and HSC/$i$-band and \textit{(ii)} their compactness in the space-based data.
Non-detections of our candidates in Hubble \textit{DASH} data argues against contamination from late-type stars.
Our constraints on the faint end of the quasar luminosity function at $z\sim6.4$ suggests a negligibly small contribution to reionization compared to the star-forming galaxy population.
The confirmation of our candidates and the evolution of number density with redshift could provide better insights into how supermassive galaxies grew in the first billion years of cosmic time.
\end{abstract}

%% Keywords should appear after the \end{abstract} command. 
%% See the online documentation for the full list of available subject
%% keywords and the rules for their use.
%\keywords{Quasars $-$ High-redshift galaxies $-$ Surveys $-$ Astronomical techniques $-$ Photometry}

%% From the front matter, we move on to the body of the paper.
%% Sections are demarcated by \section and \subsection, respectively.
%% Observe the use of the LaTeX \label
%% command after the \subsection to give a symbolic KEY to the
%% subsection for cross-referencing in a \ref command.
%% You can use LaTeX's \ref and \label commands to keep track of
%% cross-references to sections, equations, tables, and figures.
%% That way, if you change the order of any elements, LaTeX will
%% automatically renumber them.
%%
%% We recommend that authors also use the natbib \citep
%% and \citet commands to identify citations.  The citations are
%% tied to the reference list via symbolic KEYs. The KEY corresponds
%% to the KEY in the \bibitem in the reference list below. 

\section{Introduction} \label{sec:intro}

Quasi-stellar objects (QSOs) or quasars, are powered by accretion of gas on to a supermassive ($>10^{9}\,M_{\sun}$) black hole.
Since the detection of luminous quasars within the first Gyr ($z>6$) of the Big Bang using wide-area surveys such as the Sloan Digital Sky Survey, PanSTARRS, CFHTLS and VIDEO/VIKING \citep{FAN01, BECKER01, BANADOS16, MCGREER18}, it has been challenging to explain their origin and existence. Have their central engines built up their mass through sporadic, Eddington-limited accretion of gas or do they build up through more continuous accretion processes onto a relatively massive black hole seed \citep{BANADOS18, TRAKHTENBROT17}? If so, are the massive black hole seeds primordial in nature or are they end stages of early epochs of massive star-formation? Tracing the evolution of the quasar luminosity function, particularly at the low luminosity end, at the earliest cosmic times can potentially shed light on the origin of these systems. In addition, studying the number of quasars in the early Universe can help quantify their contribution to the reionization of the Universe \citep[see e.g.][]{FAN2006}.

While identification and spectroscopic confirmation of luminous $z\sim6$ quasars has become relatively straightforward, measuring the faint end ($<10^{44.5}$ erg\,s$^{-1}$, $M_{\rm UV} < -23\,{\rm mag}$) of the quasar luminosity function is much more challenging (see, for example, the comprehensive work by \citet[][]{AKIYAMA18}, \citet[][]{NIIDA20A}, and \citet[][]{MATSUOKA18} for quasar luminosity function measurements at intermediate brightness, $M_{\rm UV} \sim -23\,{\rm mag}$, at $z\sim4$, $5$, and $6$, respectively).
First, the identification of faint quasars requires a significant color difference across the Lyman-break/\lya~forest. This implies that in order to measure the break, the blue band must be at least a magnitude more sensitive. Second, faint quasars are point sources
and can lack multi-wavelength information due to the sensitivity differences between the different bands. They can be mistaken as either late-type stars in the local Universe or compact, strong emission line galaxies at intermediate redshifts.
This is especially true at seeing-limited spatial resolution of $\sim$0.7$\arcsec$. Space-resolution data with $<$0.1$\arcsec$ point spread function (PSF) full width at half maxiumum (FWHM) can help avoid this misidentification but wide-area surveys from space at such resolution are not yet available except in the COSMOS field with the {\it Hubble} Space Telescope \citep{SCOVILLE07, KOEKEMOER07}. The combined analysis of ground- and space-based data is therefore crucial to be able to robustly identify candidate, low-luminosity QSOs.

The next problem arises due to source confusion. At optical wavelengths, the classical confusion limit at seeing limited resolution is 25 AB mag (5$\times10^{44}$\,erg\,s$^{-1}$ at $z\sim6$ is 25 AB mag) as estimated from source counts in deep surveys (see Appendix~\ref{app:blending}).
From this we can estimate $\sim$30\% of objects at that brightness are contaminated by the presence of a neighboring brighter source. This makes it challenging to reliably measure colors or color limits for the sources. Disentangling the relative contributions of the confusing sources requires priors from deep, higher spatial resolution data and joint pixel level processing, taking into account accurate astrometry, and accurate position/source-dependent PSFs; this is beyond the capability of current cataloging algorithms.
Future deep, wide-area surveys conducted from space using the \textit{Euclid} and the \textit{Nancy Grace Roman} space telescopes as well as from the ground using the \textit{Vera C. Rubin} observatory are well in the regime of deep, confusion-limited, wide-area imaging. Although they promise to push the identification and characterization of \textit{luminous} quasars out to redshifts beyond $z=8$, the infrastructure for precise, joint analysis of those datasets does not yet exist.

In this paper, we demonstrate the scientific benefits of joint pixel level processing over the relatively small area of $1.64$ square degrees, to assess the faint end of the quasar luminosity function at $z\sim 6$. We combine two data sets that are very similar to the future surveys. These are the Subaru/Hyper-SuprimeCam \citep[HSC,][]{MIYASAKI18} $i$-band survey \citep{AIHARA18} and the 2004 {\it Hubble}/ACS F814W imaging survey on the roughly $2\,{\rm deg^2}$ COSMOS field \citep{SCOVILLE07, KOEKEMOER07}.
These data have similar seeing and PSF properties as future {\it Rubin} and {\it Roman/Euclid} imaging observations.

This paper is structured as follows. In Section~\ref{sec:datasets}, we present in detail the different datasets used and their preparation for extracting photometry (astrometric calibration and measurement of the PSF kernel). In Section~\ref{sec:catalog}, we detail the generation of a photometry catalog based on a joint modeling of the ACS and HSC images. Specifically, we use the high-resolution ACS images as priors to mitigate blending and confusion issues in the HSC data. In Section~\ref{sec:selection}, we present the sample selection and discuss different possible sources of contamination of our sample. We discuss our derived number density of low-luminosity quasars and impact on reionzation in Section~\ref{sec:discussion} and conclude in Section~\ref{sec:conclusions}.

Throughout this work, we assume a $\Lambda$CDM cosmology with $H_0 = 70\,{\rm km\,s^{-1}\,Mpc^{-1}}$, $\Omega_\Lambda = 0.7$, and $\Omega_{\rm m} = 0.3$. All magnitudes are given in the AB system \citep{OKE74} and stellar masses and star formation rates (SFRs) are normalised to a \citet[][]{CHABRIER03} initial mass function (IMF).

\begin{figure}[t!]
\includegraphics[angle=0,width=\columnwidth]{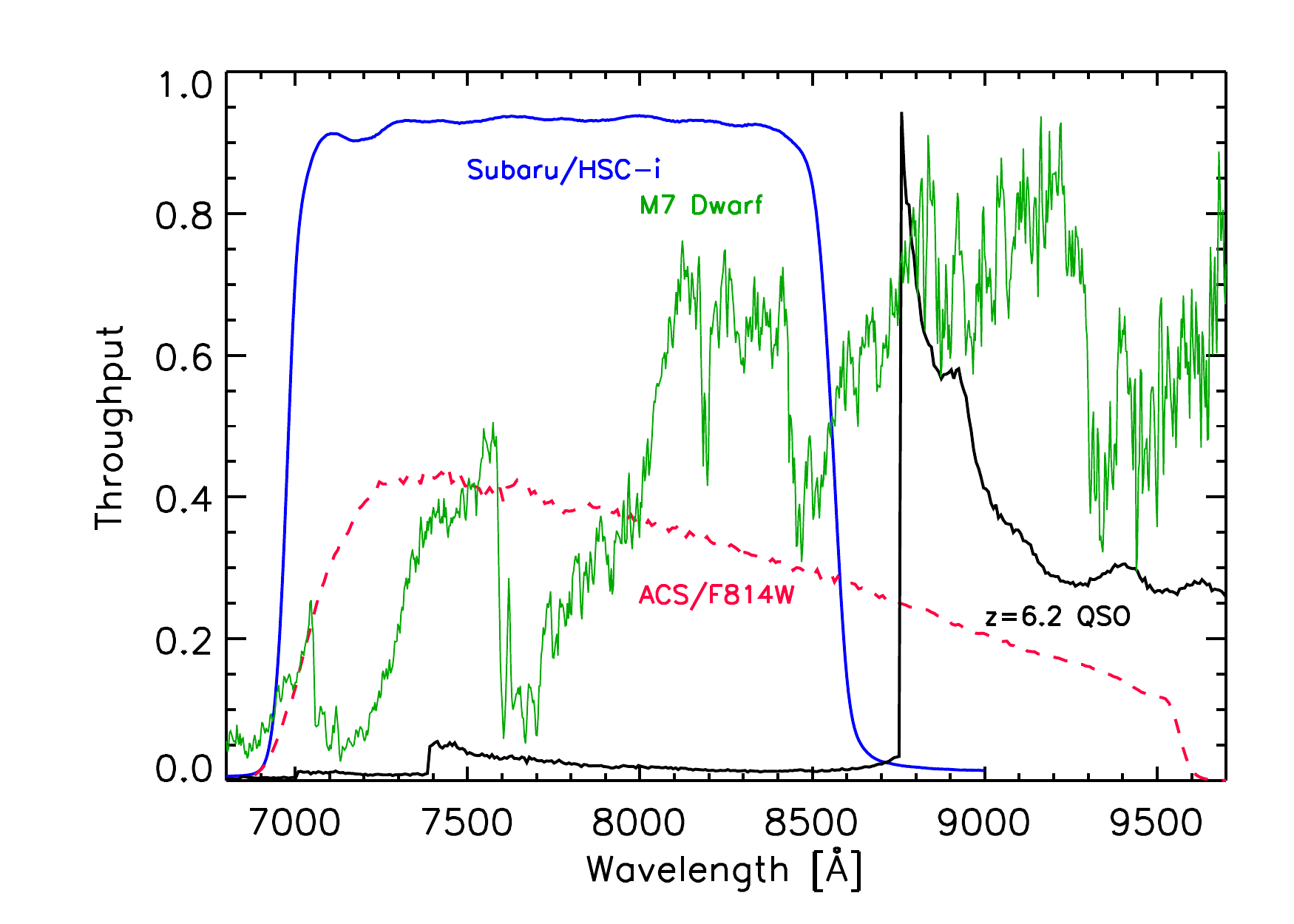}
\caption{
Throughput curves of the HSC $i$-band (blue solid) and ACS/F814W filter (red dashed). The F814W filter extends to the red, hence the flux difference acts similar to a medium-band filter. This allows the selection of quasars through a [HSC-$i$]$-$[F814W] color difference. The black line shows an average quasar template made from a stack of $\sim$2000 quasars in the Sloan Digital Sky Survey 
\citep[][]{VANDENBERK01}, redshifted to $z=6.2$ with absorption by the intergalactic medium applied \citep[][]{MADAU99}. We also show the spectrum of an M7 brown dwarf from \citet[][]{FAJARDOACOSTA16} for comparison (green).
\label{fig:filters}}
\end{figure}

\begin{figure}[t]
\includegraphics[angle=0,width=\columnwidth]{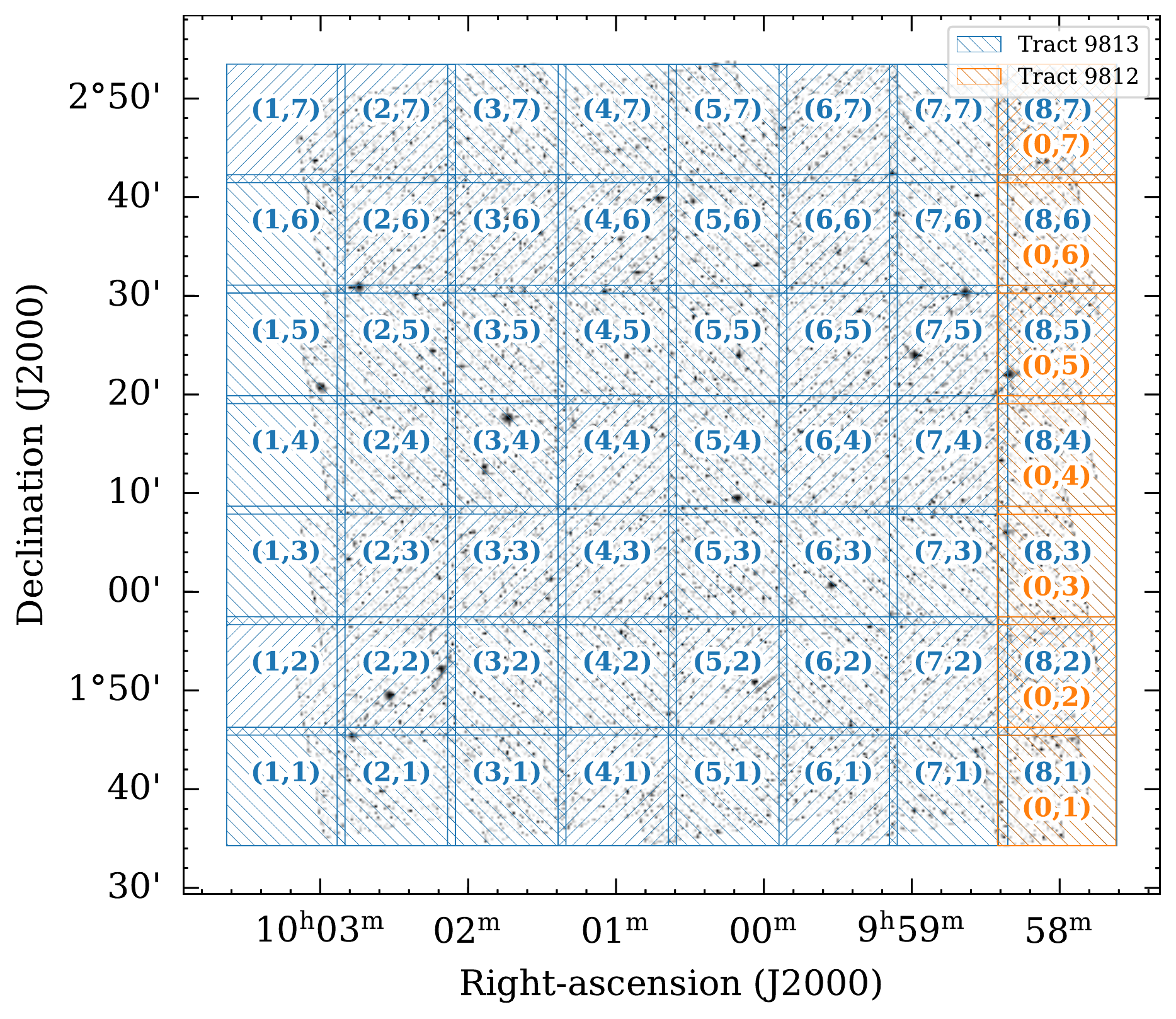}
\caption{Spatial organization of \textit{patches} on the COSMOS field. Shown are only the $63$ $12\arcmin\times12\arcmin$ \textit{patches} overlapping with the ACS/F814W observations (gray scale, background). COSMOS is covered by \textit{tract} 9813 (blue) but also has a small coverage from \textit{tract} 9812 (orange). The \textit{patch} numbers are indicated as (X,Y) pairs. Each of the \textit{patches} are divided into $16$ \textit{sub-patches} (size $3\arcmin\times3\arcmin$) to facilitate multi-processor computing. 
\label{fig:patches}}
\end{figure}

\begin{figure}[t]
\includegraphics[angle=0,width=\columnwidth]{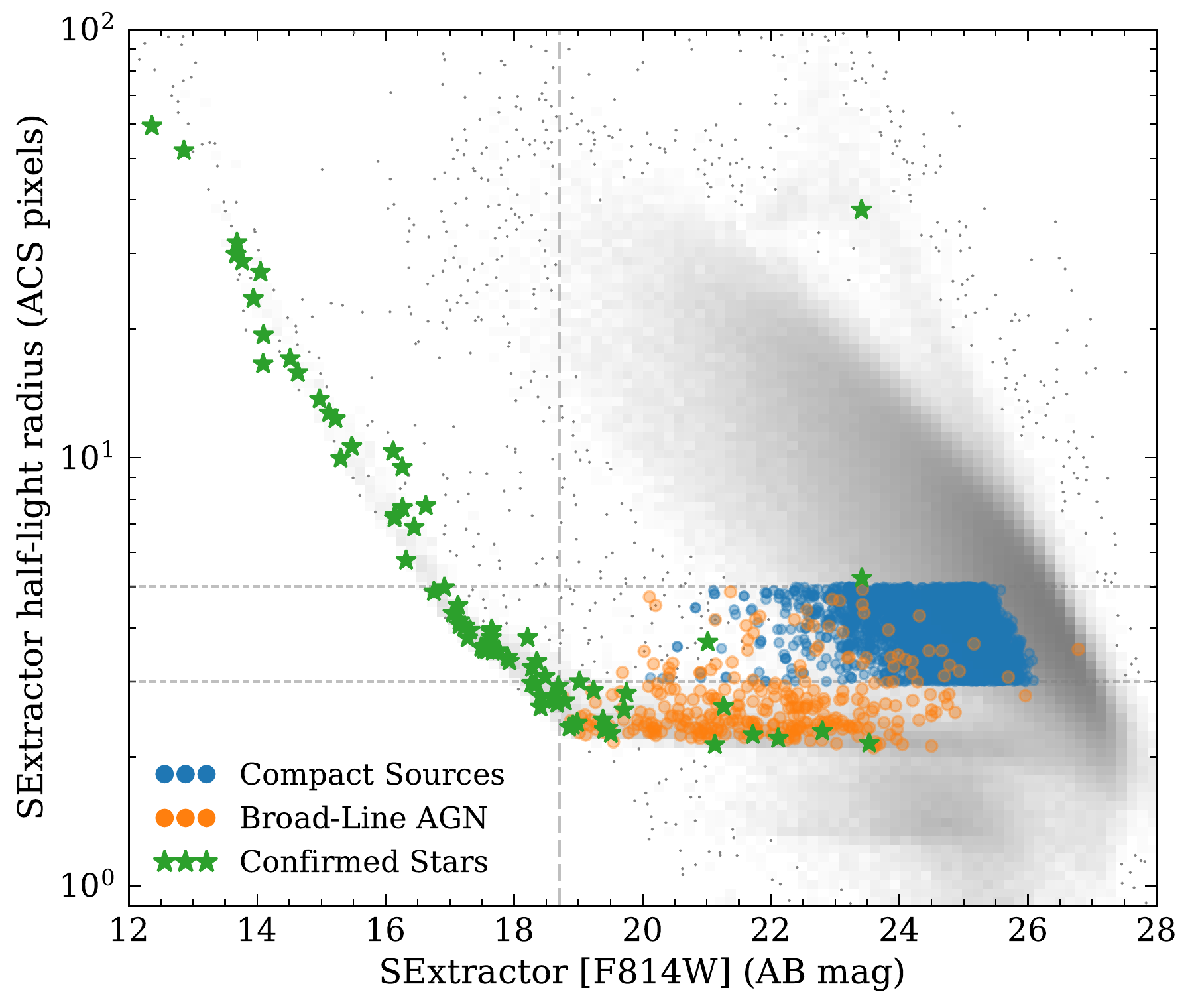}
\caption{Selection of compact sources (blue) for astrometric calibration of imaging data. Also shown are confirmed broad-line AGN \citep[orange,][]{MARCHESI16,CIVANO16,LANZUISI18} and spectroscopically confirmed stars (green). The \textit{gray} symbols (``background cloud'') are all the objects in COSMOS. The ``stellar locus,'' where point sources lie, is horizontal from right to left, and turns up around $18.7\,{\rm mag}$ (indicated by the \textit{vertical dashed line}), which is the magnitude where point sources become saturated. The horizontal \textit{dashed lines} indicate sizes of $3$ and $5$ ACS pixels. See text for more details on the selection of compact sources.
\label{fig:compactastrometryselection}}
\end{figure}

\section{Datasets and Preparation}\label{sec:datasets}

\subsection{Imaging Data}
Identification of faint $z\sim6$ QSOs requires having adequately deep imaging data for a selection similar to that of Lyman-break galaxies \citep[e.g.,][]{STEIDEL96}. Furthermore, we need at least one band to have high spatial resolution so that a separation between resolved galaxies and point-like quasars is possible.
We use the {\it Hubble}/ACS F814W images \citep{SCOVILLE07,KOEKEMOER07} and the 2018 data release 1 \citep[][]{AIHARA18} of the Hyper-SuprimeCam (HSC) $i-$band ultradeep images (called \textit{calexp}, calibrated stacked visits) from the {\it Subaru} Strategic Program (SSP) survey in COSMOS. 
As shown in Figure \ref{fig:filters}, the bandpasses and depths of these datasets are well matched, allowing for the selection of red objects between $5.9<z<6.7$.
The ACS data were taken between 2003 Oct 15 and 2005 May 21 while the HSC data were taken between 2014 March and 2015 November. Stellar proper motions result in significant astrometric mismatches between these datasets which need to be corrected for. Furthermore, these data have not been aligned to the {\it Gaia} astrometric reference frame, which needs to be remedied. 

We start with the v2.0 ACS mosaics available within the Infrared Science Archive (IRSA\footnote{\url{https://irsa.ipac.caltech.edu/Missions/cosmos.html}}), which have a PSF full width at half maximum (FWHM) of 0.095$\arcsec$ and a spatial scale of 30 mas/pixel \citep{KOEKEMOER07}. In contrast, the HSC data have a PSF FWHM between 0.6-0.9$\arcsec$ and a spatial scale of 168 mas/pixel \citep[][]{AIHARA18}. 

These data are advantageous for the search of high-redshift objects, including quasars.
The F814W bandpass overlaps entirely with the Subaru/HSC $i$-band filter and extends redward (see Figure~\ref{fig:filters} and Section~\ref{sec:basicselection}). This allows high-$z$ sources (at a median $z\sim6.4$) to be selected based on a large color difference in these bandpasses due to their Lyman break and \lya~emission. The configuration acts similarly to the selection in a medium-band filter. The sensitivity of the surveys should allow the detection of objects down to M$_{\rm UV}=-22.1\,{\rm mag}$, at least $3-4$ mags deeper than \textit{DESI} Legacy imaging surveys, over much smaller areas \citep{DEY2019}, with the added advantage of high spatial resolution in one of the bands.

%%% FIGURE ===============
\begin{figure*}[t]
\includegraphics[angle=0,width=1.11\columnwidth]{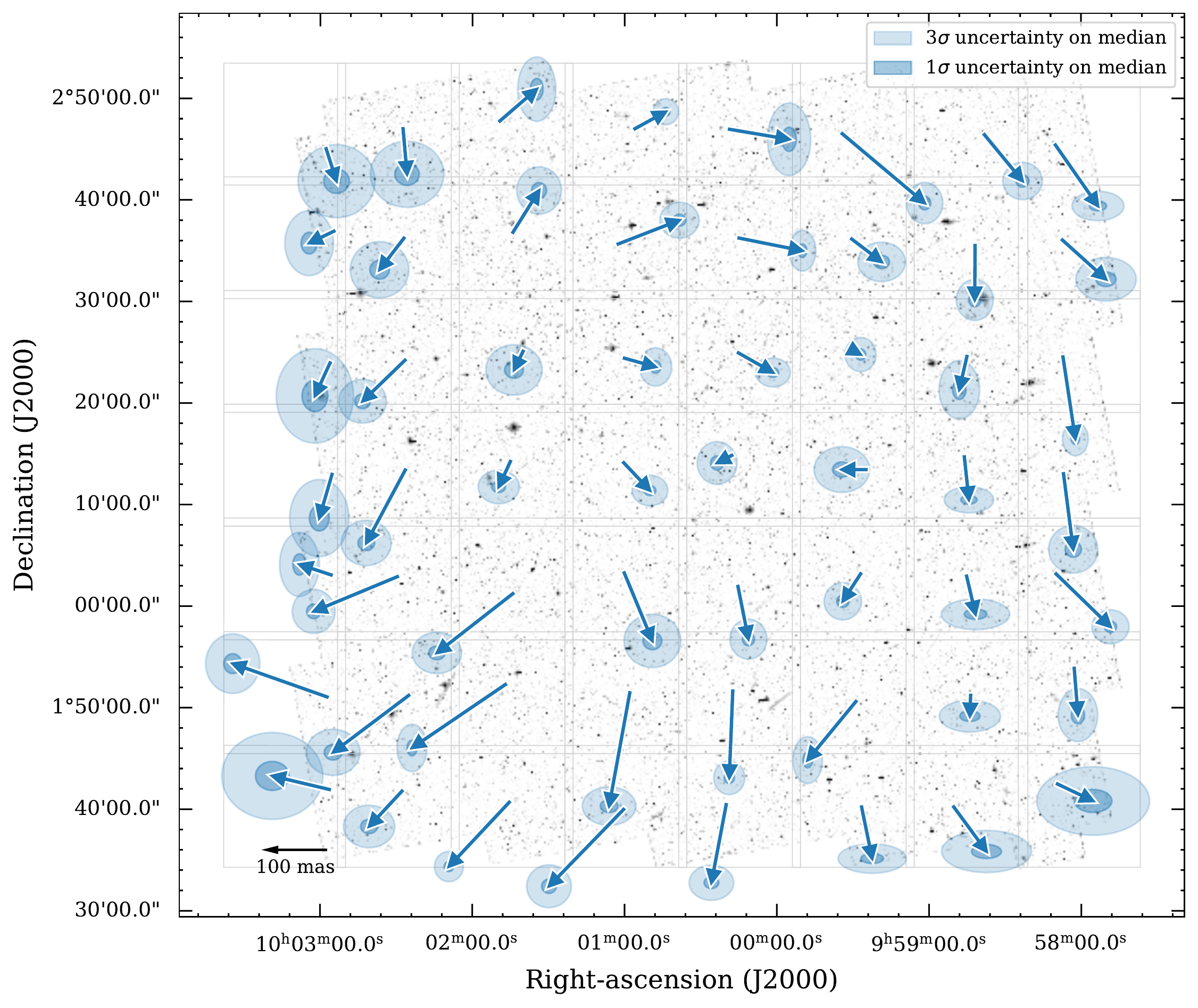}
\includegraphics[angle=0,width=0.99\columnwidth]{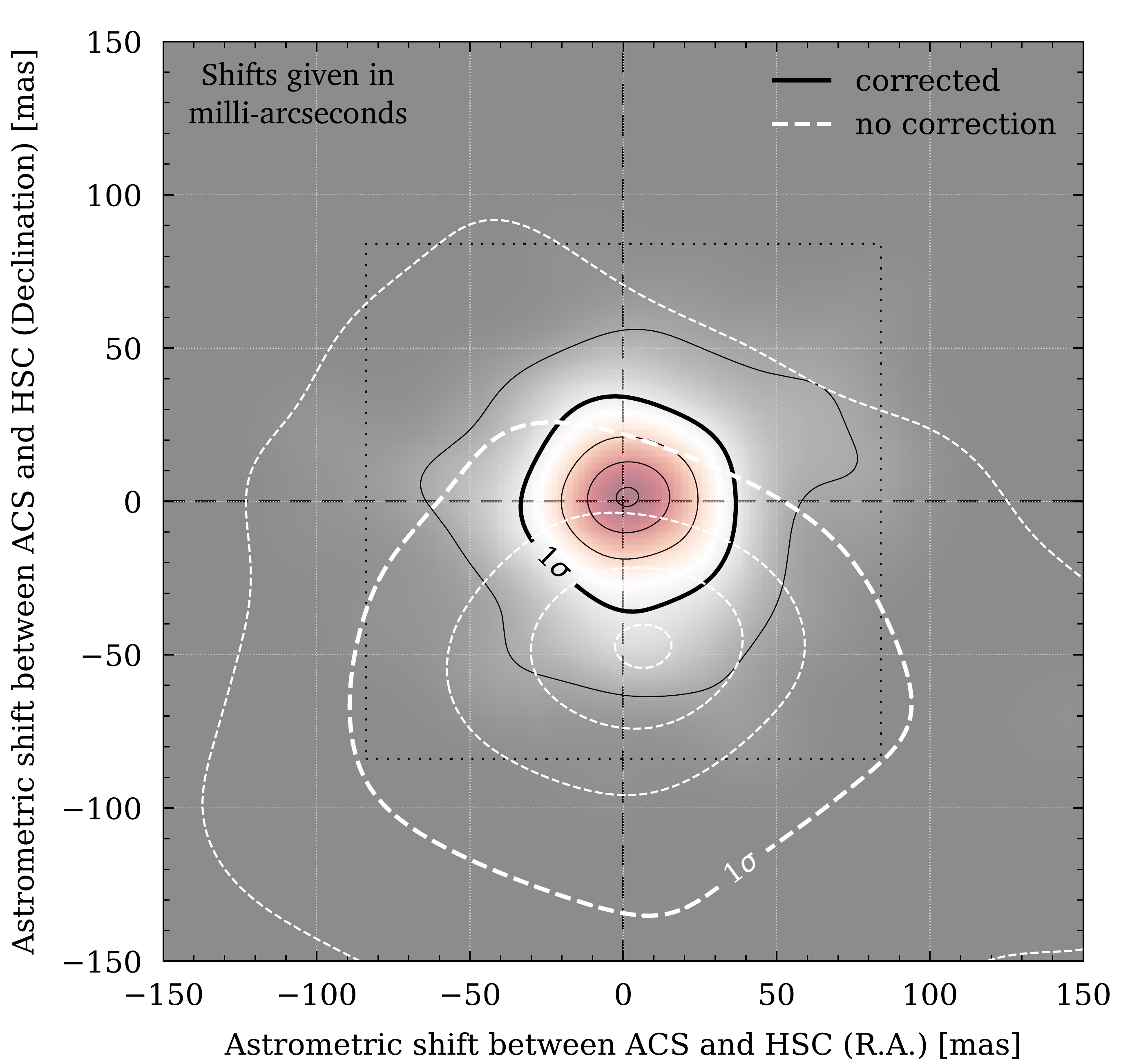}
\caption{Astrometric calibration of ACS and HSC images. \textit{Left:} Astrometric offset between ACS (from \citet[][]{KOEKEMOER07}, originally registered to the ground-based astrometric catalog from \citet[][]{CAPAK2007}) and HSC, for each of the $12\arcmin\times12\arcmin$ \textit{patches}.
The ellipses show $1\sigma$ and $3\sigma$ uncertainties in the direction and length of the vectors. The $100\,{\rm mas}$ length is shown on the lower left. \textit{Right:} Accuracy of astrometric corrections (in milli-arcseconds) tested on a sample of $150$ spectroscopically confirmed broad-line AGN \citep{MARCHESI16,CIVANO16,LANZUISI18}. The white contours show the distribution of the astrometric offsets between ACS and HSC before correction. The black contours show the offset after applying the corrections. We find no significant residual offset, and the accuracy is $\sim30-40\,{\rm mas}$.
\label{fig:astrometry}}
\end{figure*}
%%% ======================

\subsection{Data Organization Terminology}

For efficient handling and computing of joint catalogs between these datasets which are a total of $\sim$1\,TB, we organized the data in \textit{tracts}, \textit{patches}, and \textit{sub-patches}.
The HSC data is organized in $1.7\,{\rm deg} \times 1.7\,{\rm deg}$ \textit{tracts}. Each of them is split in 81 $12\arcmin\times12\arcmin$ \textit{patches} \citep[see][]{AIHARA18}. The \textit{tracts} have a $1\arcmin$ overlapping region, while the \textit{patches} overlap within $100\,{\rm px}$, which corresponds to $16.8\arcsec$. The \textit{patches} themselves are stacks of different visits. 

The ACS observations on the COSMOS field are covered by \textit{tract} numbers $9812$ and $9813$, including $63$ \textit{patches} of the total $133$ \textit{patches} contained in the ultra-deep data. To allow multi-processing of these data in the following, we cut the \textit{patches} into $16$ \textit{sub-patches} of size $3\arcmin\times3\arcmin$ and overlap of $10\arcsec$. Hence, in total there are $1008$ \textit{sub-patches} to be processed (Figure~\ref{fig:patches}).

While the HSC images are already cut into \textit{patches}, we scripted a Python wrapper to use the online IRSA cutout tool\footnote{\url{https://irsa.ipac.caltech.edu/data/COSMOS/index_cutouts.html}} to cut and retrieve the ACS images at the correct size and location from the IRSA server. The \textit{sub-patches} are then created using the \texttt{Cutout2D} Python package provided by \textit{Astropy}\footnote{\url{http://www.astropy.org}} \citep[][]{ASTROPY13,ASTROPY18}.

\subsection{Astrometric Calibration of Imaging Data}\label{sec:astrometry}

The alignment of images (i.e., their relative astrometric calibration) is crucial to perform any successful joint pixel level processing. 
Among others, an accurate astrometric calibration of the images allows us to use the $\sim11$ year time baseline between ACS and HSC data to study the proper motions of faint stars \citep[see][]{FAJARDOACOSTA21}. For this work, an accurate relative astrometric calibration is needed to use the priors on location and sizes from the {\it Hubble}/ACS imaging, to mitigate the effects of blending and confusion in ground based images (see Section~\ref{sec:catalog}). The HSC data used here had been aligned with PanSTARRS DR1, while the astrometric reference frame for the ACS data was defined by CFHT $i-$band mosaics \citep{CAPAK2007} that covered the full COSMOS field to a 5$\sigma$ limiting depth of $26.2\,{\rm mag}$. It provided $\sim$ 300-600 sources on each ACS tile, that could be used for relative registration. Each ACS tile was then registered to the CFHT grid, and the overall relative alignment precision was found to be $\sim$ 5-10 milliarcseconds \citep[see also][]{KOEKEMOER11}. However, this relative precision between the ACS and the CFHT sources does not account for possible absolute errors in the CFHT astrometry, which we explore.

We use two methods to align the ACS and HSC images \textit{patch} by \textit{patch}.
In the first method, we compute the absolute astrometric alignment of the images using 3937 stars from the \textit{Gaia} DR2 catalog \citep{LANZUISI18}, matched to ACS sources within a 0.2 arcsec radius. Since Gaia stars are generally bright, saturation is an issue especially for the deep ground and space-based images used here. 
To explore the saturation limit, we run \textsc{SExtractor} \citep[][]{BERTIN96} on the ACS images to generate a catalog with approximate source sizes (half-light radii).
We define a ``stellar locus" in a magnitude versus size plot as shown in Figure~\ref{fig:compactastrometryselection}. Typically, stars would be unresolved and would have FWHM sizes of $\sim$2-3 ACS pixels. However, bright/saturated stars have a larger fraction of their PSF profile above the noise threshold which is the reason for their size increasing as a function of brightness. 
From the upturn of the ``stellar locus'' on this diagram, as well as Gaussian fits to individual stars, we estimate a saturation threshold of about $18.7\,{\rm mag}$. The faintest Gaia stars in our ACS sample were $20.9\,{\rm mag}$ in our catalog and so we use sources in the magnitude range $18.7-20.9$ {\rm mag} which have less than 3 pixels half light radius, to estimate the astrometry.
We identified the positions of these objects on the HSC images using 2-dimensional Gaussian fits. This was not necessary for the ACS images
because the centroiding accuracy is much smaller than the
typical astrometric scatter that we expect.
For each of the different \textit{patches}, we then computed the true position of the \textit{Gaia} stars from their DR2 catalog proper motion. While the epochs of the HSC images are within 2015 (close to the epoch of \textit{Gaia}, 2015.5), the ACS images were taken between 2003 and 2005. We therefore computed the mean epoch of each ACS source from the individual exposures that covered it. For HSC, all sources within a \textit{patch} were measured simultaneously. The mean of the offsets between true and measured positions of the \textit{Gaia} stars are then used to compute the astrometric offset between ACS and HSC of a given \textit{patch}.

For the second method, we use compact extra-galactic sources on both images to compute their relative astrometric offset (Figure~\ref{fig:compactastrometryselection}). Ideally, one would use quasars or AGN for this, as they are point sources and do not have proper motions. However, the density of bright quasars is less than a few per square degree. Instead, we select our compact sources to have \textit{(i)} a minor-to-major (B/A) axis ratio of more than 0.9 on the ACS images, \textit{(ii)} a signal-to-noise (S/N) ratio of more than 25 in ACS, \textit{(iii)} an ACS magnitude fainter than $20\,{\rm mag}$, and \textit{(iv)} sizes between 3 and 5 ACS pixels (corresponding to $0.09\arcsec-0.15\arcsec$). In addition, we avoid blended sources by restricting the sample to those with \textsc{SExtractor} flag \texttt{FLAG}$=0$ (on ACS and HSC images). Finally, we apply a S/N threshold of $10$ for their HSC photometry measured by \textsc{SExtractor}.
In total, we select about $5000$ compact sources. Their position on the HSC and ACS images are then compared to compute the astrometric offset between these images.

We find that both methods lead to similar astrometric offset corrections (within $\sim3-5\,{\rm mas}$) and therefore choose to use the second method in what follows.
Furthermore, we note that some of the bright Gaia stars could be in the non-linear/saturation regime even at $20\,{\rm mag}$. This could potentially effect the astrometric correction. However, since we find almost identical astrometric offsets using Gaia stars and faint compact sources, we think that this effect is negligible.

The left panel of Figure~\ref{fig:astrometry} shows the average relative astrometric offset between ACS and HSC images per $12\arcmin\times12\arcmin$ \textit{patch}. The ellipses show the $1\sigma$ and $3\sigma$ uncertainties in direction, and the length of a $100\,{\rm mas}$ shift is indicated on the lower left. Note the clear clock-wise circular pattern that could be due to inhomogeneous astrometric calibration due to distortions on either the ACS or HSC images. Subsequent data releases both by HSC and a re-reduction of the ACS data (G. Brammer et al., private communication) have fixed these offsets but at the time this work was initiated, these astrometric offsets were present.

The right panel of Figure~\ref{fig:astrometry} is an assessment of the accuracy of our astrometric calibration. Specifically, we apply the astrometric shifts to a sample of $150$ spectroscopically confirmed broad-line AGN \citep[see Figure~\ref{fig:compactastrometryselection}; ][]{MARCHESI16,CIVANO16,LANZUISI18}. These are point sources and do not have proper motions, hence are ideal to test the astrometric calibration. The white contours show the shift in R.A. and Declination before applying the astrometric correction between HSC and ACS. The black contours show the distribution after correction. It is centered at zero (as it should be because AGN do not have motions) with a $1\sigma$ width of $30-40\,{\rm mas}$. The latter is our precision of astrometric calibration.
For more details, we refer to our companion paper on proper motions of faint stars by \citet[][]{FAJARDOACOSTA21}. We anticipate that in space-based data with larger fields of view resulting in higher source numbers, the astrometric precision can be improved further.

%%% FIGURE ===============
\begin{figure}[t]
%\vspace{5cm}
\includegraphics[angle=0,width=\columnwidth]{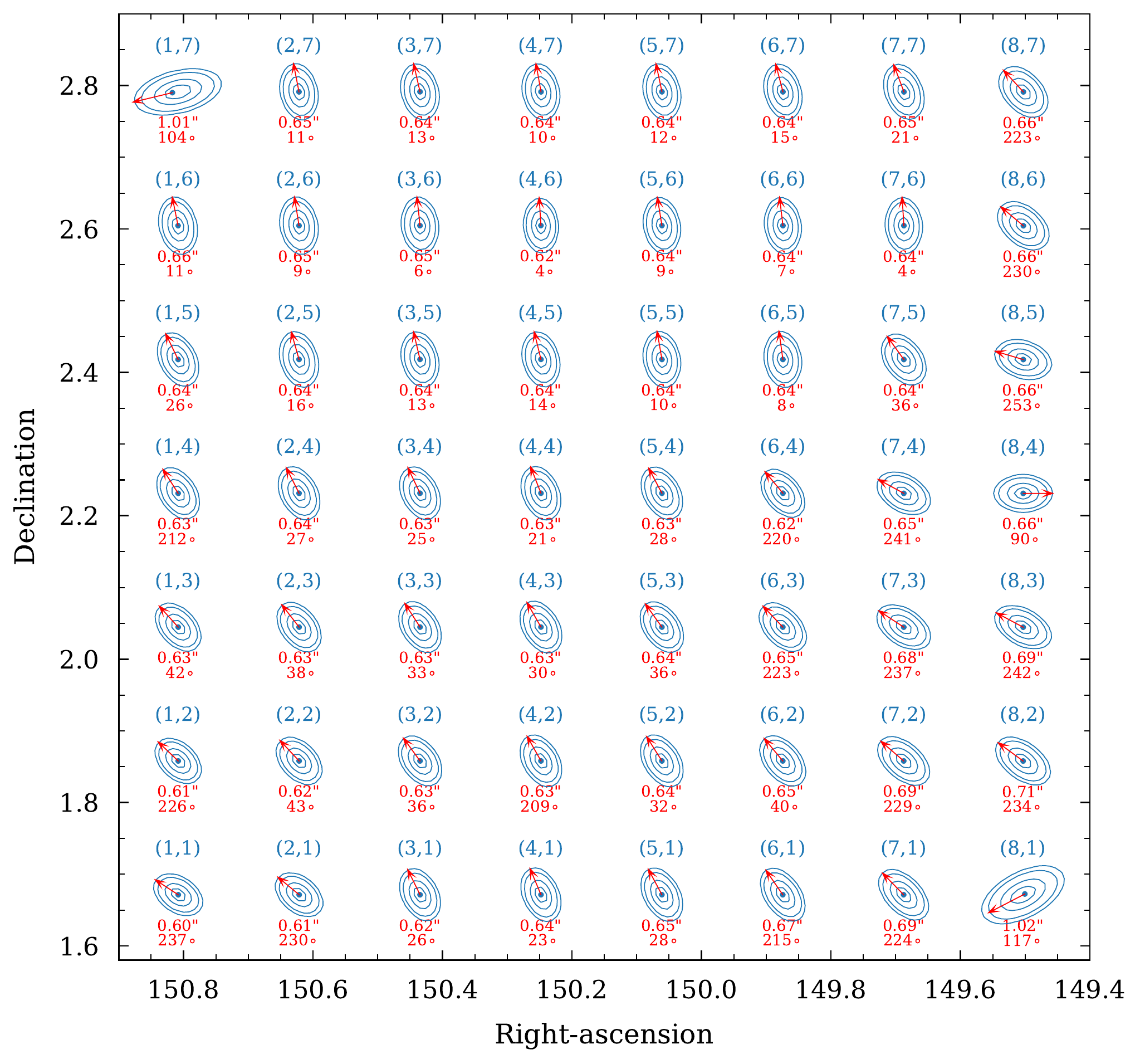}
\caption{Variations in the HSC PSF across the COSMOS field. Generally, the variations are on a very small level, however we notice that the PSFs have a preferred north-south direction (ellipticity exagerated by $50\%$). The contours show the 20\%, 50\%, 80\%, and 90\% enclosed flux of the PSF. The red arrows show the semi-major axis (with FWHM and position angle indicated). The \textit{patch} numbers are indicated in blue.
\label{fig:psfvariations}}
\end{figure}
%%% ======================

\subsection{PSF estimation}\label{sec:psf}

In addition to the astrometry, the PSF of both HSC and ACS images has to be known accurately in order to produce reliable photometry. To create a spatially varying PSF, we stack unsaturated \textit{Gaia} stars as well as fainter stars. We select the latter on the ACS images by extrapolating the ``stellar locus'', fit by the \textit{Gaia} stars on the magnitude vs. size diagram (Figure \ref{fig:compactastrometryselection}), to lower magnitudes of $23\,{\rm mag}$. To create the stacks, we use the code \texttt{PSFex} \citep{BERTIN11}, which creates a magnitude-dependent model PSF based on a linear combination of (sub-pixel centered) stars in a given catalog. It also outputs the residual after subtracting a scaled PSF for each of the stars in the sample, which is useful to access the quality of the fits. We found that the residuals are smaller by a few percent for a magnitude dependent fit compared to a simple stack. Depending on the ACS coverage, we are able to use between $30$ and $100$ stars per $12\arcmin\times12\arcmin$ \textit{patch} for the fit. The ACS PSF FWHM is less than $0.1\arcsec$, while the HSC PSF FWHM varies between $0.60\arcsec$ and $0.75\arcsec$ with a median around $0.64\arcsec$. We find that the HSC PSFs have a preferred direction, which is approximately north-south. Other variations in rotation and ellipticity between the different \textit{patches} are relatively minor but nonetheless have to be taken into account to measure robust photometry (Figure~\ref{fig:psfvariations}).

%%% FIGURE ===============
\begin{figure*}[t]
\includegraphics[angle=0,width=2.1\columnwidth]{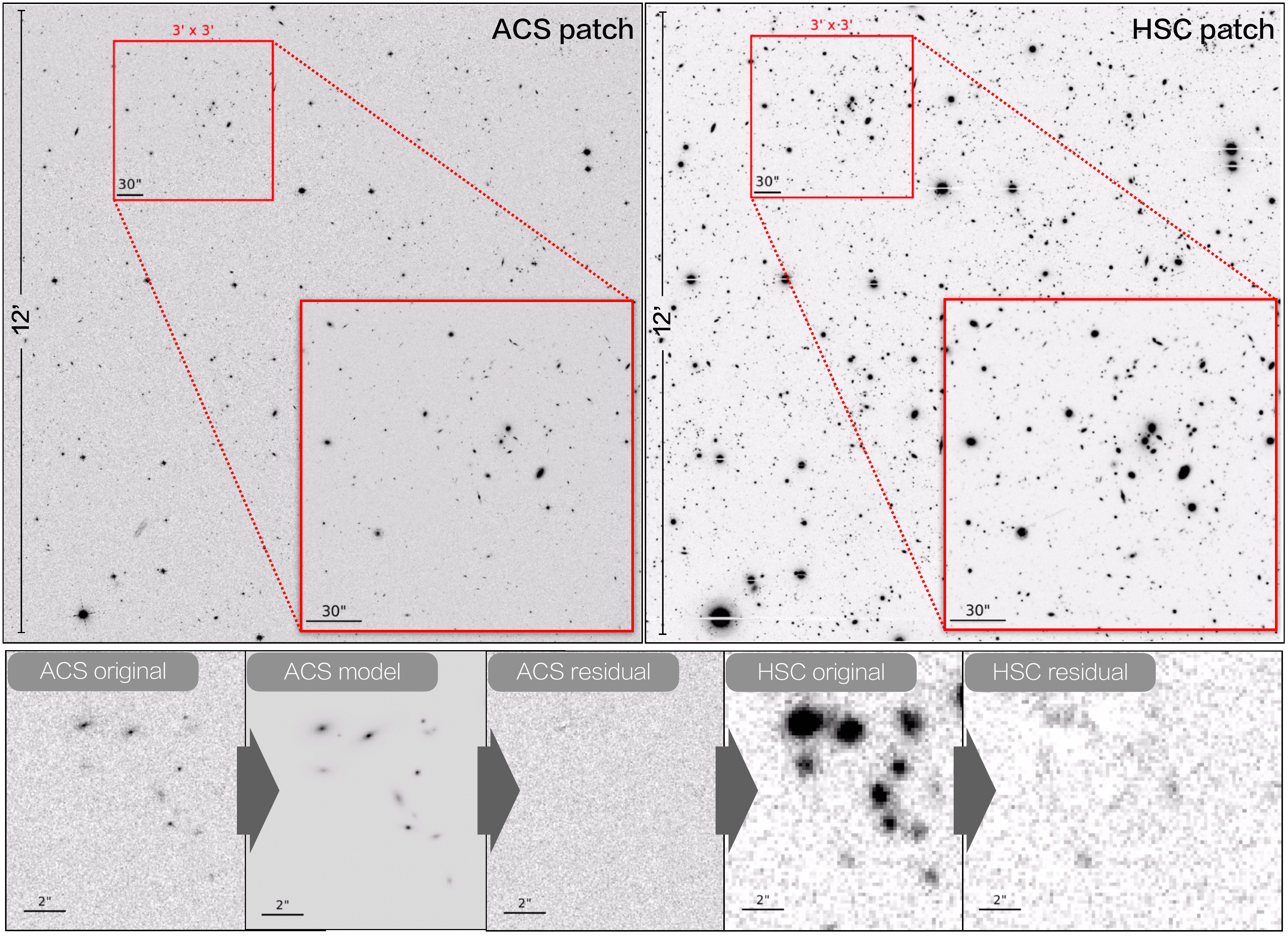}
\caption{Illustration of our prior-based photometery approach with \textsc{Tractor}. The top large panels show the $12\arcmin\times12\arcmin$ \textit{patches} of the ACS and HSC observations. The inset shows a $3\arcmin\times3\arcmin$ region, which is the size of a \textit{sub-patch}. The bottom row of panels shows different stages of our approach. From left to right these are \textit{(i)} original ACS image, \textit{(ii)} ACS model image created with \textsc{Tractor}, \textit{(iii)} ACS residual image (original minus model), \textit{(iv)} HSC original image, and \textit{(v)} HSC residual image after convolving the ACS models with the difference in the PSFs and scaling to the HSC images of the sources. This prior based approach has the advantage that it can efficiently deblend sources when applied to deep, low-resolution images.
\label{fig:tractorillustration}}
\end{figure*}
%%% ======================

\section{Joint Cataloging with \textsc{Tractor}} \label{sec:catalog}

Once the astrometric offset between the ground- and space-based datasets is established, and the PSFs kernels for each dataset are calculated, we can undertake joint pixel-level photometry. This takes into account the location and morphological extent of the sources in the space-resolution data to alleviate the role of source confusion and accurately measure the photometry in the ground-based data. Here we use the code \textsc{Tractor}\footnote{\url{http://thetractor.org/}} \citep{LANG16b,LANG16a,WEAVER21}. 

In brief, \textsc{Tractor} performs a parametric shape fit to a source in an image by using a maximum likelihood analysis including a weight map of the image. Thereby multiple sources can be fit simultaneously on an image, which is the preferred way to run \textsc{Tractor} to obtain robust photometric measurements \citep[see also detailed description and testing in][]{WEAVER21b}.

\textsc{Tractor} has two major advantages compared to classical photometry codes such as \textsc{SExtractor} or other aperture-based methods. First, it can use shape priors derived from a high-resolution image to photometer low-resolution images. To do so, \textsc{Tractor} can be forced to fix the shape and position and only vary the normalization (i.e., total flux) to minimize the residuals. In overlapping bands such as HSC$-i$ and ACS F814W the morphological $k-$correction is negligible, hence this is a reasonable assumption.
Since \textsc{Tractor} can fit nearby sources of light simultaneously, this forced-photometry approach is valuable in fields with high confusion or blending. 
Second, because of the parametric fitting approach and the inclusion of the PSF (in the model step), \textsc{Tractor} provides PSF-corrected total fluxes for the fitted objects; this is especially valuable for extracting the contribution of low surface brightness regions beyond the isophotal size of objects. 

On the other hand, there are regimes where the performance of \textsc{Tractor} is significantly reduced. Although \textsc{Tractor} provides a variety of models to parameterize the light distribution, including point sources (i.e., $\delta-$functions), simple Gaussians, Sersic models, and even multi-Gaussian representations, it would fail to measure robust fluxes of non-smooth light distributions. Examples are galaxies with extended spiral structure or bulge and disk components (specifically when fit with a single Sersic profile), or irregular galaxies such as lumpy galaxies at high redshifts.
As we later demonstrate through simulations (Section~\ref{sec:simulations}), we found that for our scientific goal (measurement of the photometry of compact sources), \textsc{Tractor} results in accurate photometry.

In Appendix~\ref{app:tphot} we compare the performance of our method to another extensively used photometry package called \textsc{TPhot} \citep{MERLIN15,MERLIN16}, which is currently being upgraded in preparation for \textit{Euclid} joint photometry. We find very good agreement between both methods.

\subsection{ACS and HSC Photometer Pipeline} \label{sec:pipeline}

Our pipeline takes as input an ACS and HSC \textit{patch} together with the corresponding PSF (Section~\ref{sec:psf}) and astrometric shifts between the images (Section~\ref{sec:astrometry}). It then runs \textsc{Tractor} on the ACS image to create a parametric model for each source in the image. The model for each source is then convolved by the (positional dependent) HSC PSF and scaled to fit the data in the HSC image (see Figure~\ref{fig:tractorillustration}). We detail the different steps in the following. Multiple \textit{sub-patches} are processed in parallel (see Appendix~\ref{app:nersc}).

\textbf{Step 1 - Preparation.} We first cut a $3\arcmin\times3\arcmin$ \textit{sub-patch} with $10\arcsec$ overlap from the HSC \textit{patch} together with the corresponding ACS \textit{sub-patch}. The overlap ensures a good fitting of sources at the edges. To keep the World Coordinate System (WCS) information, we are using the Python command \texttt{Cutout2D} from the \textit{AstroPy} package. The following steps are applied to a single \textit{sub-patch}.

\textbf{Step 2 - Obtain positions and preliminary shapes.} We first run \textsc{SExtractor} on the ACS image. This provides the locations of the sources as well as initial shape parameters (such as \texttt{A\_IMAGE}, \texttt{B\_IMAGE}, \texttt{FLUX\_AUTO}, \texttt{FLUX\_RADIUS}, and \texttt{THETA\_IMAGE}) which are needed as initial guesses for \textsc{Tractor}. It also produces a segmentation map which identifies the pixels out to the isophotal radius of each source.
The ACS data has very few repeats per pixel on the sky as a result of which there are residual cosmic rays in the mosaic.
We perform a removal of such spurious sources in the SExtractor catalog, which are characterized by sizes smaller than the diffraction limit, with \texttt{FLUX\_RADIUS} $<1.7$ ACS pixels. We found that this cut removes most cosmic rays and spurious sources clustered at the edges of the ACS coverage.

\textbf{Step 3 - Obtain ACS models and photometry.} \textsc{Tractor} is run on each source extracted in step 2. Specifically, we create a cutout for each source using the \textsc{SExtractor} \texttt{XMIN}, \texttt{XMAX}, \texttt{YMIN}, and \texttt{YMAX} keywords. We increase the size of this cutout by 50 percent and fit all objects within this cutout (also the ones not covered entirely) simultaneously.
It is important to distinguish between unresolved and resolved sources for obtaining the best possible fits. Specifically, unresolved sources (or point sources) are fit using \textsc{Tractor}'s \textit{PointSource}\footnote{The \textit{PointSource} class requires a position and total flux.} class. All other sources are fit using the \textit{SersicGalaxy}\footnote{The \textit{SersicGalaxy} class requires a position, total flux, axis ratio, position angle, half-light radius, and Sersic index $n$.} class. We assign a point source flag to all sources with either \textit{(i)} \texttt{CLASS\_STAR} greater than $0.8$, brighter than $23\,{\rm mag}$, and axis ratio\footnote{In the following, defined as the ratio of \texttt{B\_IMAGE} and \texttt{A\_IMAGE}.} greater than $0.8$, or, \textit{(ii)} fainter than $23\,{\rm mag}$, axis ratio greater than $0.8$, and half-light radius smaller than the F814W PSF FWHM. The parameters measured by \textsc{SExtractor} in step 2 and a Sersic index $n=2$ are used as initial guesses. The \textsc{Tractor} \textit{Image} object is created using the \textit{PixelizedPSF} class, which converts the input PSF from \textit{FITS} format to a format suitable for \textsc{Tractor}. Furthermore, we feed in the per-pixel noise measured from a $3\sigma$ clipping on the image. The background is fixed at zero level as the images are background subtracted.
During fitting, we allow the position to wander within $\pm1\,{\rm px}$ (corresponding to $\pm30\,{\rm mas}$) from its initial guess. All other parameters, except the position angle (given by \texttt{THETA\_IMAGE}), are set free to vary.

\textbf{Step 4 - Forced photometry on the HSC image.} Finally, the best-fit parameters (position, flux, and shape parameters in the case of resolved sources) obtained in step 3 are used to photometer the HSC image. In a similar manner as in the previous step, cutouts of each source are created. However, as we expect significant blending and confusion compared to the ACS image, we cannot simply take the segmentation image created in step 2. Instead, we first run \textsc{SExtractor} on the HSC image and enlarge the resulting segmentation map by convolving it with a $1\arcsec$ boxcar filter. This will combine the segmentation areas of close-by objects, creating one single large area\footnote{Note that the resulting segmentation map is a binary map as the knowledge of pixel ownership for the extracted sources is lost in the convolution step.}. From this, we recalculate the \texttt{XMIN}, \texttt{XMAX}, \texttt{YMIN}, and \texttt{YMAX}. For each source in the ACS \textsc{Tractor} catalog (from step 3), we check in which enlarged segmentation area it falls and create a cutout accordingly. All sources within this cutout are fit simultaneously using their ACS position and shape priors. If no segmentation area was found (e.g,. due to faintness or color gradient), we apply a $2\arcsec\times2\arcsec$ cutout size.
Note that we add the astrometric offsets between the HSC and ACS images obtained in Section~\ref{sec:astrometry} to the ACS prior position in this step.
During the fit, we allow the positions to vary within $\pm0.5\,{\rm px}$ (corresponding to $\pm 84\,{\rm mas}$) and fix all other parameters except the total flux. 
We note that because we are applying a median astrometric offset between the ACS and HSC images per \textit{patch}, allowing the positions to vary a small amount is crucial to account for astrometric scatter.

\begin{figure}[t!]
\includegraphics[angle=0,width=1.0\columnwidth]{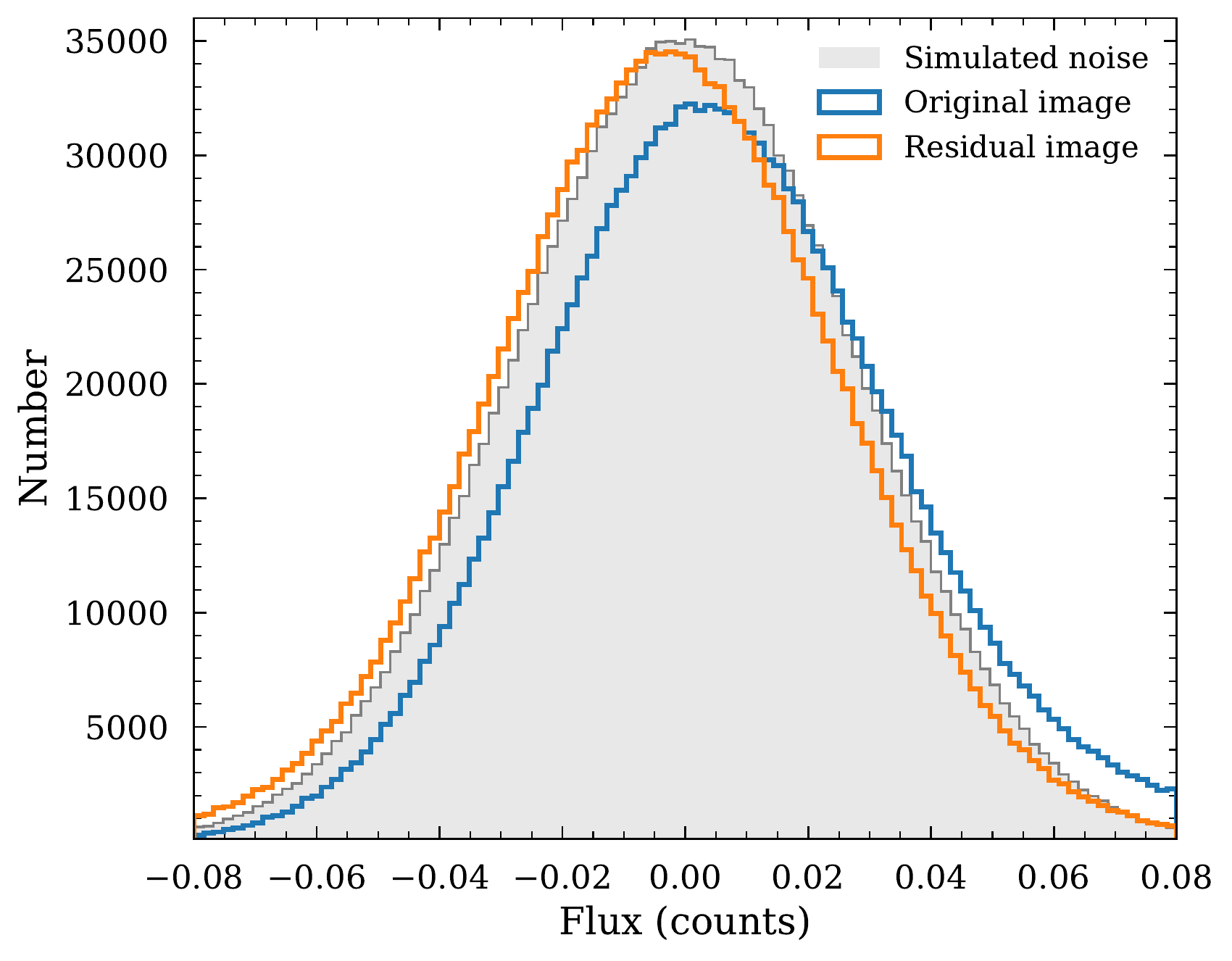}
\caption{Comparison of the pixel flux distribution on the original HSC image (blue) and the residual HSC image after photometering with \textsc{Tractor} (orange). The gray histogram shows the expected distribution for uniform noise across the image. The peak of the residuals is shifted slightly negative which indicates a slight ``over subtraction'' of sources (i.e., over estimation of flux).
\label{fig:residual}}
\end{figure}

Figure~\ref{fig:residual} shows the histograms of the pixel flux distribution on the original HSC image as well as the residual image after fitting all sources with \textsc{Tractor}. While the original image (blue line) shows a tail towards positive fluxes (these are the detected sources), the residual image (orange line) is in good agreement with the expected noise distribution (simulated uniform Gaussian noise across image without sources). There is a small imbalance on the residual image towards negative fluxes, which could hint towards a slight over-subtraction of fluxes by \textsc{Tractor}. Statistically, this offset is less than $0.01\,{\rm mag}$ at $25.5\,{\rm mag}$.

%%% FIGURE ===============
%%% ======================

Appendix~\ref{app:nersc} provides a detailed descriptions of the setup and the process to run the pipeline efficiently at the \textit{National Energy Research Scientific Computing Center} (NERSC).

\subsection{Simulations}\label{sec:simulations}

To demonstrate the performance of our pipeline in measuring point source fluxes and to compute the sensitivity limits and detection completeness, we created simulated images including points sources of different brightness using the software \textsc{SkyMaker}\footnote{\url{https://www.astromatic.net/software/skymaker}} \citep{BERTIN09}. In total, we simulate $15,000$ point sources in the magnitude range $24-27\,{\rm mag}$ and $30,000$ point sources in the range $27-29\,{\rm mag}$ arranged in a grid to avoid blending (we discuss the effects of blending in Appendix~\ref{app:blending}). The increased number of faint sources provides us with a more uniform distribution of sources across this magnitude range. The pixel noise properties (specifically the per-pixel RMS measured in empty apertures) of the simulated images match the ones of the real ACS and HSC images. We found that correlated noise is negligible and is therefore not included. We also use the measured PSF of ACS and HSC. The generated simulated images are then run through our pipeline with the same setup as used for the real images.

\begin{figure}[t!]
\includegraphics[angle=0,width=1.0\columnwidth]{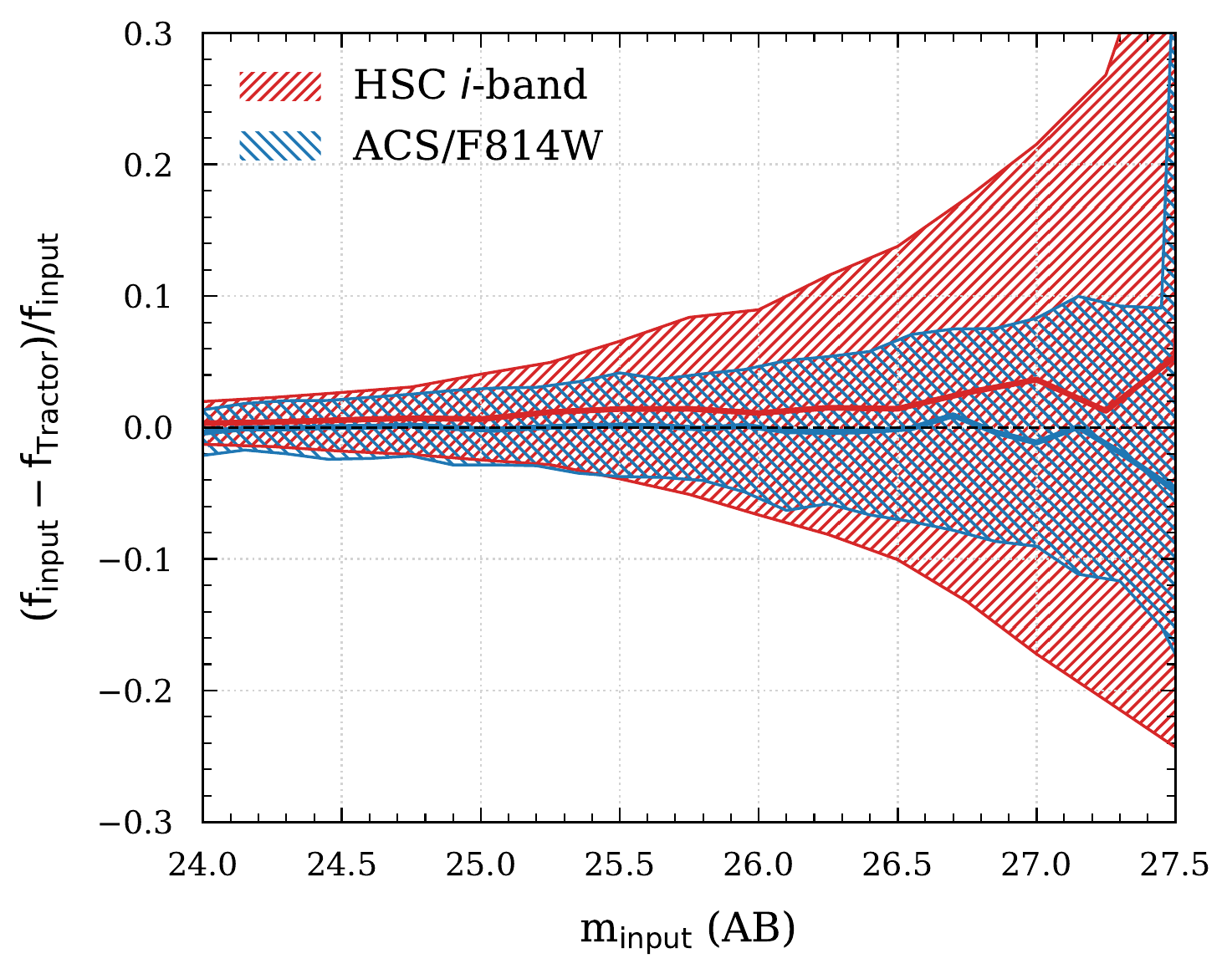}
\caption{Recovery of point source photometry as a function of magnitude for ACS (blue) and HSC (red). The hatched regions show the 16$^{\rm th}$/84$^{\rm th}$ percentiles of the scatter and the solid lines show the running median. Undetected point sources are included with $f_{\rm tractor} = 0$. We do not find any significant biases in the point source photometry.
\label{fig:accuracy}}
\end{figure}

Figure~\ref{fig:accuracy} shows the difference between input and recovered magnitudes for F814W (blue) and HSC $i-$band (red) as a result of this test. The median (solid lines) and $68\%$ percentiles (hatched regions) are computed in running bins of size $0.2\,{\rm mag}$.
As expected, the scatter increases towards fainter magnitudes. The $1\sigma$ width of the distribution (here computed as the width around the median that contains $68\%$ of the points) in each magnitude bin reaches $0.21\,{\rm mag} = 2.5/\ln(10)/5\,{\rm mag}$ (corresponding to $5\sigma$) at $26.9\,{\rm mag}$ for HSC $i-$band and $27.4\,{\rm mag}$ for F814W, respectively. These are our obtained $5\sigma$ sensitivities. The latter is consistent with the number quoted in \citet[][]{KOEKEMOER07}. Note that the scatter includes undetected point sources, for which we set their flux value to zero. Generally, we find biases of less than $0.02\,{\rm mag}$ in the measured fluxes of compact sources down to the $5\sigma$ sensitivity limit of the HSC $i-$band.

As we photometer the HSC images using an ACS prior, the detection completeness of our catalog depends on the detection on the ACS image entirely. The latter is performed by \textsc{SExtractor}. It is therefore straightforward to compute the overall detection completeness for point sources. Using the same simulated image as above, we find a detection completeness of $100\%$ down to the HSC $5\sigma$ limit and a decrease to $90\%$ at $27.5\,{\rm mag}$ in F814W.

%%% FIGURE ===============
%%% ======================

\section{Selection of Quasars Candidates}\label{sec:selection}

Quasars at high redshift can be identified primarily by their compactness and color. By definition, quasars should be point-like and therefore unresolved, even in space-based observations. Furthermore, the ACS F814W filter extends redward of the Subaru/HSC $i$-band filter, which allows the measurement of the flux difference across rest-frame wavelengths of $1216\,{\rm \AA}$ (Figure~\ref{fig:filters}). For quasars (or any high redshift galaxy) a red color is expected due to the drop in flux caused by the \lya~forest \citep[absorption of ionizing photons by intervening neutral Hydrogen, see, e.g.,][]{STEIDEL96,FAN01}.

In the following, we describe the selection of our high-redshift quasar candidates in more detail.

\begin{figure}[t!]
\includegraphics[angle=0,width=\columnwidth]{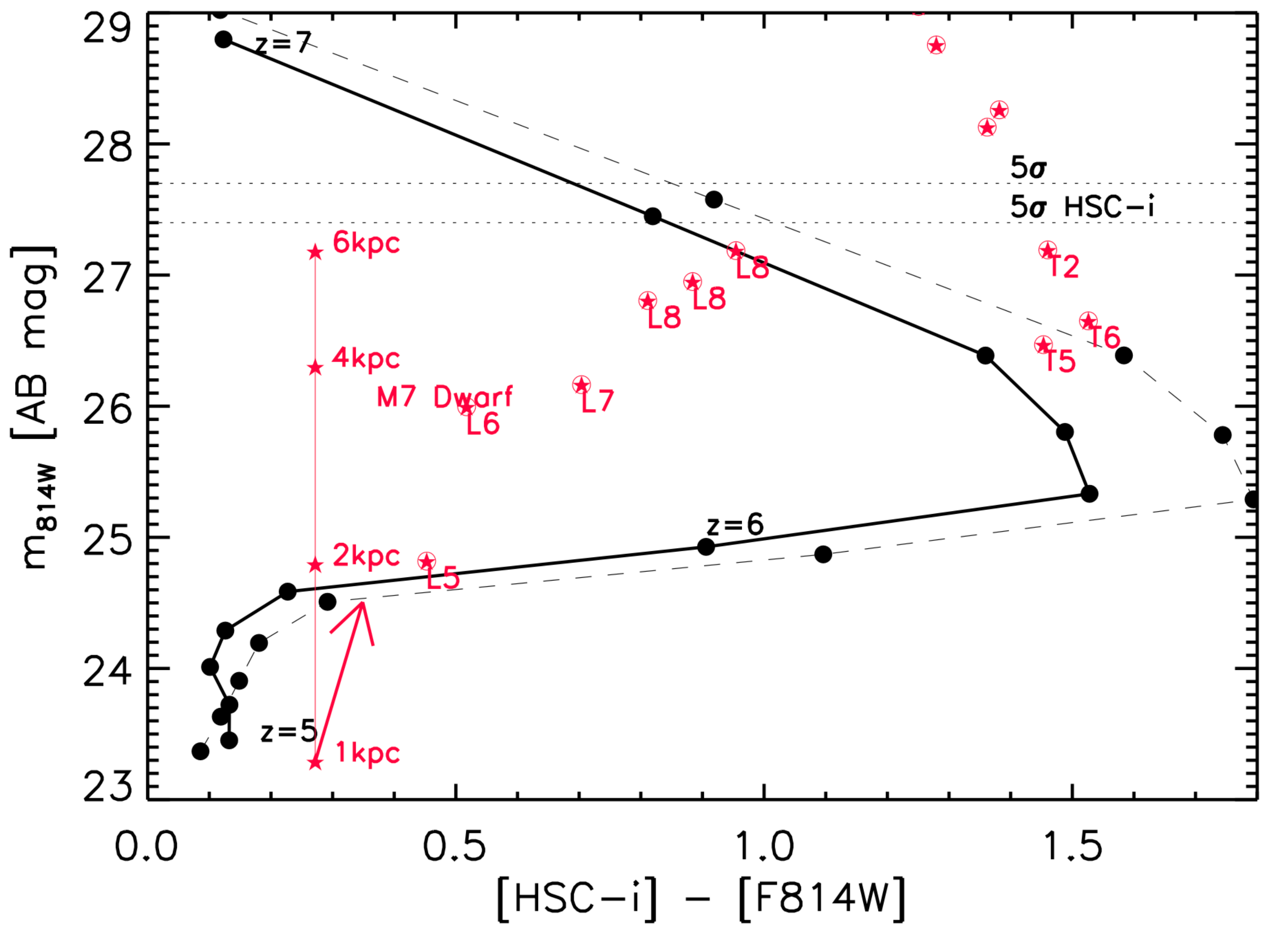}
\caption{
High-redshift objects at $z>6$ can be selected through a color difference between photometry in the {\it Subaru}/Hyper-SuprimeCam $i-$band and the {\it Hubble}/ACS F814W filter (see also Figure~\ref{fig:filters}). The lines show the expected [HSC-i]$-$[F814W] color for an average quasar template at redshifts $z=5-7$ and normalized to a $B-{\rm band}$ luminosity of $10^{45}$\,erg\,s$^{-1}$ (black line and dots). The black dots are separated by $\Delta z = 0.2$. The solid black lines shows the template from \citet[][]{TELFER02}, while the dashed line shows the template from \citet[][]{VANDENBERK01}.
For galaxies and quasars at $5.9<z<6.8$, the [HSC-i]$-$[F814W] colors will be $>0.5\,{\rm mag}$ (for both templates used) mainly due to absorption by the Lyman-$\alpha$ forest blueward of $1215\,{\rm \AA}$ as well as the \lya~emission line.
The red symbols denote different spectral types of brown dwarfs at their native distances from \url{DwarfArchives.org}. The red line shows the M7 dwarf from Figure 1 at distances between $1-6\,{\rm kpc}$. 
The impact of a A$_{V}=2\,{\rm mag}$ extinction on the stellar spectra is shown by the red arrow. We discuss the contamination by stars in Section~\ref{sec:contamin_stars} in more detail.
\label{fig:colorselection}}
\end{figure}

\subsection{Selection by Color and Compactness}\label{sec:basicselection}

First, we perform a selection of high-z quasars from the joint catalogs, based on color and compactness.

As shown in Figure~\ref{fig:filters}, the combination of the HSC $i-$band and the ACS F814W filter allows the selection of high-redshift objects by the [HSC-i]$-$[F814W] color. 

Figure~\ref{fig:colorselection} shows the expected [HSC-i]$-$[F814W] colors for the quasar template in Figure~\ref{fig:filters} as well as a template from \citet[][]{TELFER02} at different redshifts, normalized to a rest-frame $B-{\rm band}$ luminosity of $10^{45}\,{\rm erg\,s^{-1}}$ and with a \citet[][]{MADAU99} IGM absorption applied. We tested more realistic realizations of IGM absorption \citep[e.g.,][]{INOUE14} and found changes of less than $0.01\,{\rm mag}$ in color, which would not change the results in the following.
Note that luminosity (an unknown parameter here) does not affect the color, only the value on the $y$-axis.

We therefore apply an initial [HSC-i]$-$[F814W] color cut of $>0.5\,{\rm mag}$, which selects objects between redshifts $5.9 \leq z \leq 6.8$. Note that this selection also removes stars of spectral types earlier than L5 (even dust obscured, see Figure~\ref{fig:colorselection}), which are spatially unresolved and therefore would pass the compactness criteria. The contamination by cooler stars will be discussed in Section~\ref{sec:contamin_stars}.
For the color selection, we include the uncertainties and limits of the magnitude measurement as well as a $3\sigma$ clearance. Specifically, we require
\begin{equation}\label{eq:colorcut}
    m_{\rm HSC} - m_{\rm ACS} > 0.5 + 3\times\sqrt{\sigma_{\rm HSC}^2 + \sigma_{\rm ACS}^2},
\end{equation}
where $\sigma_{\rm HSC}$ and $\sigma_{\rm ACS}$ are the $1\sigma$ uncertainties on the HSC-$i$ and F814W magnitudes, respectively. For HSC-detected sources, we take the $1\sigma$ flux errors output by \textsc{Tractor}. For sources detected at less than a S/N of $5$, we assume the $1\sigma$ limit measured from our simulations (Section~\ref{sec:simulations}) for a S/N$=5$ source. Note that galaxies or quasars at $z\geq6$ should generally not be detected in the HSC $i$-band (they are so-called $i$-band dropouts). To be most inclusive in our initial selection, we include sources detected in HSC-$i$, which will later be removed in our final visual inspection (Section~\ref{sec:visualinspection}).

%%% FIGURE ===============
\begin{figure}[t!]
\includegraphics[angle=0,width=1.0\columnwidth]{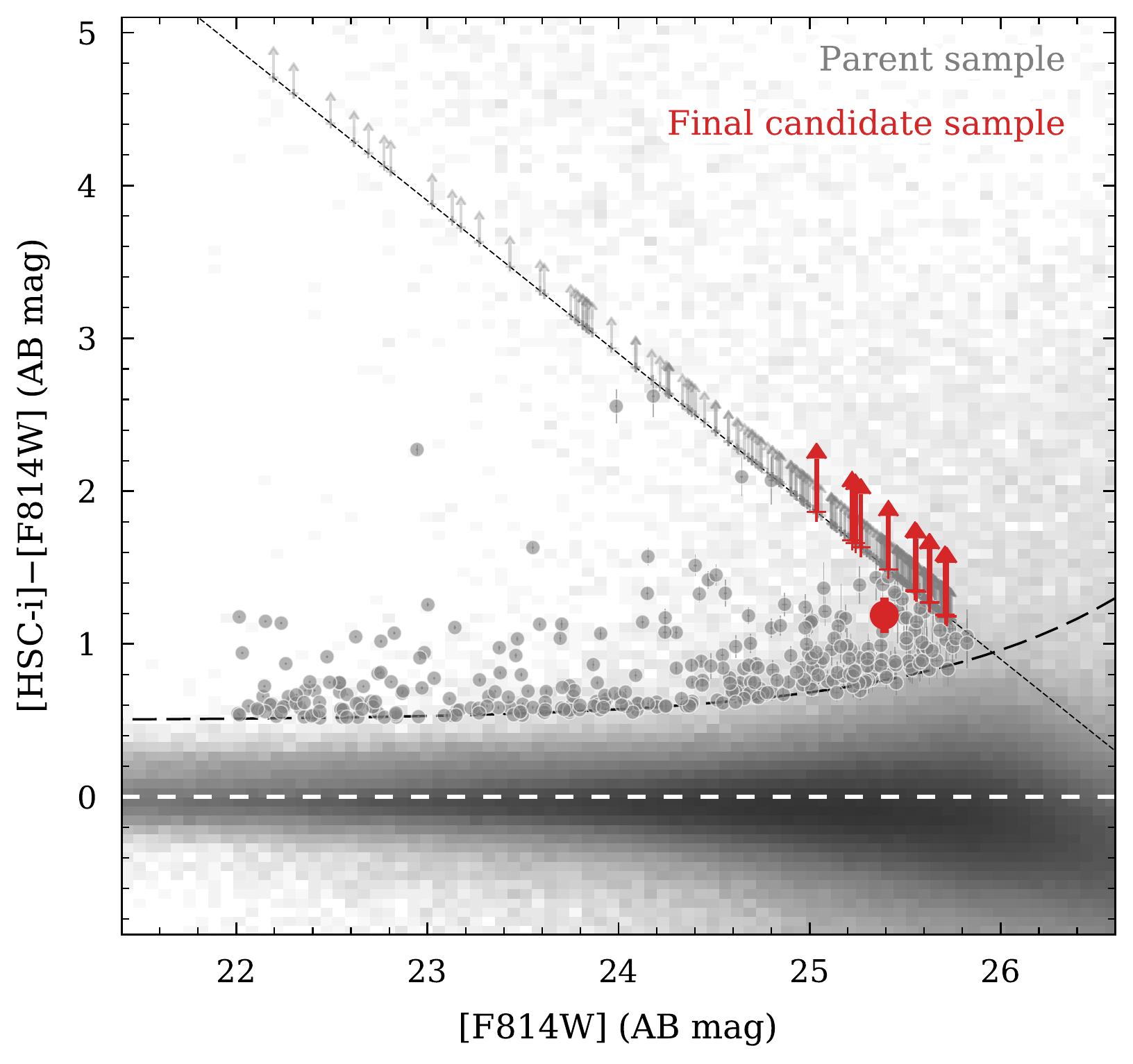}
\caption{Initial [HSC-i]$-$[F814W] color selection. The gray background cloud shows all extracted sources in our \textsc{Tractor} catalog, while the gray symbols show the $555$ galaxies from the initial selection by color and compactness. The points denote $>5\sigma$ detections in HSC-$i$, while the arrows show upper limits. The final sample of $12$ candidates is indicated in red.
The long-dashed line shows Equation~\ref{eq:colorcut}, assuming the theoretical relation between magnitude uncertainty and $1\sigma$ limit. The dotted line shows the $5\sigma$ point-source limit in HSC-$i$.
\label{fig:finalselection}}
\end{figure}
%%% ======================

LBGs, as well as strong emission line galaxies at lower redshifts (such as \oiii~at $0.72<z<0.88$), will have similar colors and therefore could be included in this selection. We estimate that the latter would result in [HSC-i]-[F814W] colors of $\sim0.2$ or less, hence would be excluded by our color cut.
Furthermore, LBGs and low-$z$ galaxies would be more spatially extended than quasars given the high resolution of the HST observations (see Section~\ref{sec:contamin_lbg}).
We therefore require an additional cut in ACS size. 

Specifically, we select sources that are compact and unresolved in the F814W images. In the following, we use the \textsc{SExtractor}-derived \texttt{CLASS\_STAR}, \texttt{FLUX\_RADIUS} (half-light radius), and axis ratio to perform this selection.
From our point source simulations described in Section~\ref{sec:simulations}, we find that \textsc{SExtractor}'s star classification neural network can reliably tag point sources with \texttt{CLASS\_STAR} $>0.9$ down to a magnitude of $25.5\,{\rm mag}$. For fainter magnitudes, \texttt{CLASS\_STAR} $>0.5$ includes point sources, but with a significant amount of contamination from extended sources. We therefore add an additional \texttt{FLUX\_RADIUS} cut of $3$ ACS pixels, which corresponds to the average PSF FWHM. In addition, we require an axis ratio (\texttt{B\_IMAGE}/\texttt{A\_IMAGE}) of $>0.8$, which, according to our point source simulations, is expected for point sources brighter than $27\,{\rm mag}$ in F814W.

After applying these two selection criteria, we remove sources at the edges of the ACS coverage, which likely are spurious due to the reduced number of frames causing lower sensitivities and less reliable cosmic ray/artifact removal.
We end up with a sample of 555 sources, which are shown in Figure~\ref{fig:finalselection}, along with the other objects in our \textsc{Tractor} catalog.

\subsection{Visual Inspection}\label{sec:visualinspection}

Next, we visually inspect the ACS and HSC images and the corresponding \textsc{Tractor} residuals of the $555$ candidates. The main purpose is to remove obvious stars from our sample and to ensure a non-detection in HSC-$i$ and bluer bands (see below). The latter is expected for quasars and galaxies at $z>6$. 

\begin{figure}[t]
\includegraphics[angle=0,width=1\columnwidth]{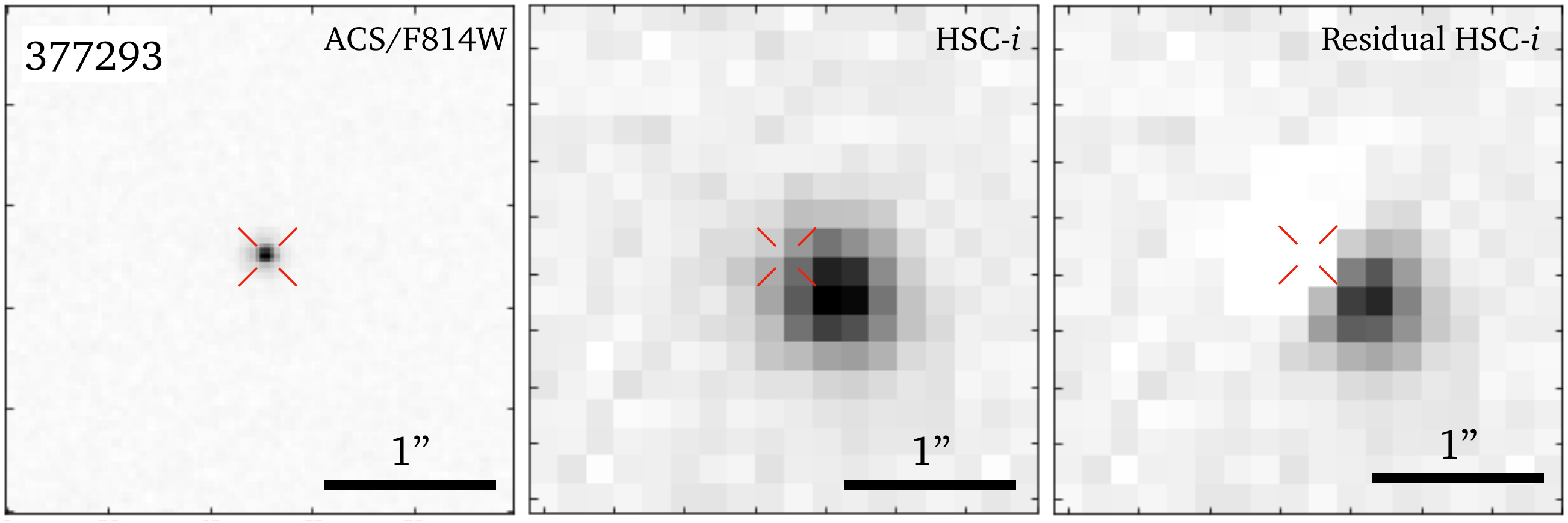}
\caption{Example of a $23\,{\rm mag}$ star with significant proper motion ($\sim15\,{\rm mas\,yr^{-1}}$ measured between 2004 and 2015). This star was included in our selection due to the under-estimation of the HSC-$i$ flux by \textsc{Tractor}. The latter is introduced by semi-fixed prior positions.
\label{fig:propermotion}}
\end{figure}

We notice that some stars with proper motion are included in our candidate selection due to an underestimated flux measurement in HSC, hence resulting in a red color. This is because we let the prior position centroid only vary by $\pm0.5$ HSC pixels (corresponding to about $80\,{\rm mas}$) to improve deblending, hence stars with proper motions of more than $\sim8\,{\rm mas\,yr^{-1}}$ ($80\,{\rm mas}$ over the 10 year baseline of ACS and HSC) are expected to be fit poorly by \textsc{Tractor}. Figure~\ref{fig:propermotion} visualizes this in the extreme case of a $23\,{\rm mag}$ star with a proper motion of roughly $\sim15\,{\rm mas\,yr^{-1}}$.

%%% FIGURE ===============
%%% ======================

To ensure a non-detection in HSC-$i$ and bluer bands, we created stacks of the ancillary data in various broad-band and narrow-band filters, which we downloaded directly from IRSA.
Specifically, we create two stacks; a ``blue'' stack including the Subaru $B$, $V$, $r$, $g$, and $NB816$ (narrow-band) images  \citep[][]{TANIGUCHI07,TANIGUCHI15}, and a ``red'' stack consisting of the Subaru $z-{\rm band}$ and the UltraVISTA $Y$, $J$, $H$, and $K-{\rm band}$ images  \citep[][]{MCCRACKEN12}\footnote{The $3\sigma$ point source sensitivities in magnitudes (taken from \citet[][]{LAIGLE16} and \citet[][]{CAPAK2007}) are $27.6$, $26.9$, $27.0$, $27.0$, $25.8$, $26.4$, $25.8$, $25.4$, $25.0$, $25.2$ for the $B$, $V$, $r$, $g$, $NB816$, $z$, $Y$, $J$, $H$, and $K-$bands, respectively}.
If the sources are truly at high redshifts, the blue stack should result in a non-detection. We therefore exclude candidates that have a visual detection in the blue stack as well as the HSC-$i$ filter. On the other hand, a detection in the red stack is possible and could serve as confirmation that the source is real, despite the significantly lower sensitivity in those bands. For example, a $z=6.0$ quasar with $\sim25.5\,{\rm mag}$ in F814W would be expected to have a magnitude of $\sim25$ in the UltraVISTA filters and could be marginally detected in the stack. For this calculation, we used the same quasar templates as in Figure~\ref{fig:colorselection}, which we normalized accordingly and then convolved with the UltraVISTA filters. We therefore do not impose any selection criteria on the red stack.

Finally, we also remove some of the obvious spurious sources (such as objects on the edges of the field or diffraction spikes of stars) and candidates with unreliable photometry where the contribution from a bright foreground source could not be accurately removed.

After this selection step, we end up with $33$ candidates that obey the color cut (Equation~\ref{eq:colorcut}), are compact, and are undetected in the ``blue" stack.

\subsection{Removal of Spurious Sources (Cosmic Rays)}\label{sec:removalspurious}

\begin{figure}[t]
\includegraphics[angle=0,width=1\columnwidth]{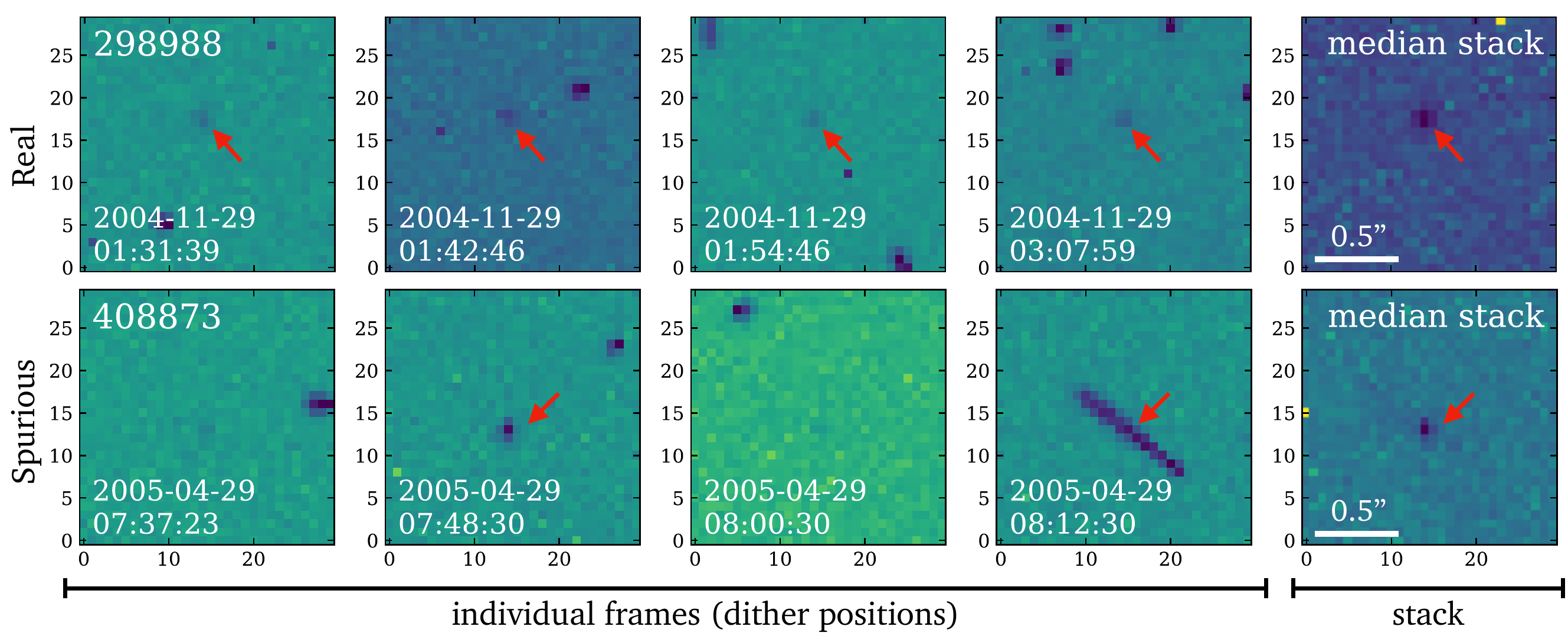}
\caption{Example of a real source (top row) and a spurious source (cosmic ray, bottom row) on images taken in the ACS/F814W filter. The left four panels show the individual frames (i.e., dither positions, date and time of observation are indicated). The right-most panel shows a simple median stack of the individual frames. The red arrow marks detections. We remove sources from our candidate sample if they are detected in fewer than three frames.
\label{fig:example_spurious}}
\end{figure}

Our best candidates from above are only detected in ACS/F814W and therefore may include spurious artifacts in the HST images. Specifically, due to the preference of surveying a large area, each ACS/F814W pointing on COSMOS only used a 4-point dither pattern, making cosmic ray removal more challenging. To alleviate this, we require that candidates are detected in at least 3 frames (out of 4). Note that due to the superior depth of the ACS images compared to HSC, our candidates are detected at ${\rm S/N}\gtrsim30$ in F814W, hence should be detected in individual frames as well at a ${\rm S/N}\gtrsim15$.
Figure~\ref{fig:example_spurious} shows an example of a real detection and a spurious detection. The latter is likely caused by cosmic ray hits on frames two and four, while the former is at consistent flux levels in all four dither positions.
After the visual inspection and rejection of spurious detections, we end up with $12$ final candidates in our sample.

%%% FIGURE ===============
%%% ======================

%%% FIGURE ===============
\begin{figure}[t]
\includegraphics[angle=0,width=1\columnwidth]{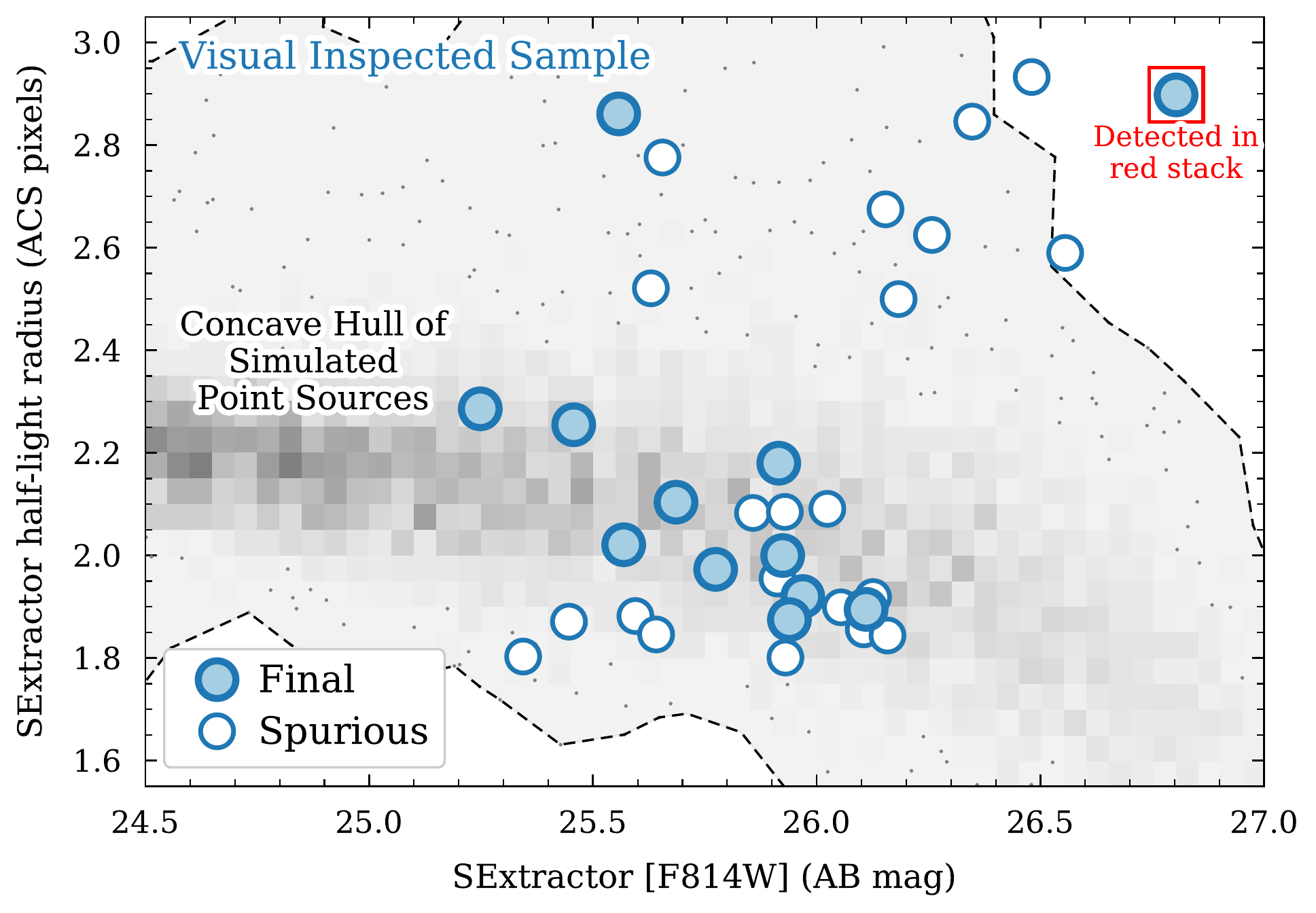}
\caption{Visually inspected sample of $33$ candidates on the \texttt{MAG\_AUTO} vs. \texttt{FLUX\_RADIUS} diagram. The final $12$ candidates are shown in filled blue circles. Candidate $772319$, detected in the red stack and with $z_{\rm phot} = 5.92$, is highlighted with a red box. Visually identified spurious detections (Section~\ref{sec:removalspurious}) are shown as empty blue circles. The envelope of all simulated point sources (see text) is shown in gray. It includes our final sample of candidates thus suggesting that they are real.
\label{fig:mag_vs_size}}
\end{figure}
%%% ======================

%%% FIGURE ===============
\begin{figure*}[t]
\begin{center}
    \includegraphics[angle=0,width=1.3\columnwidth]{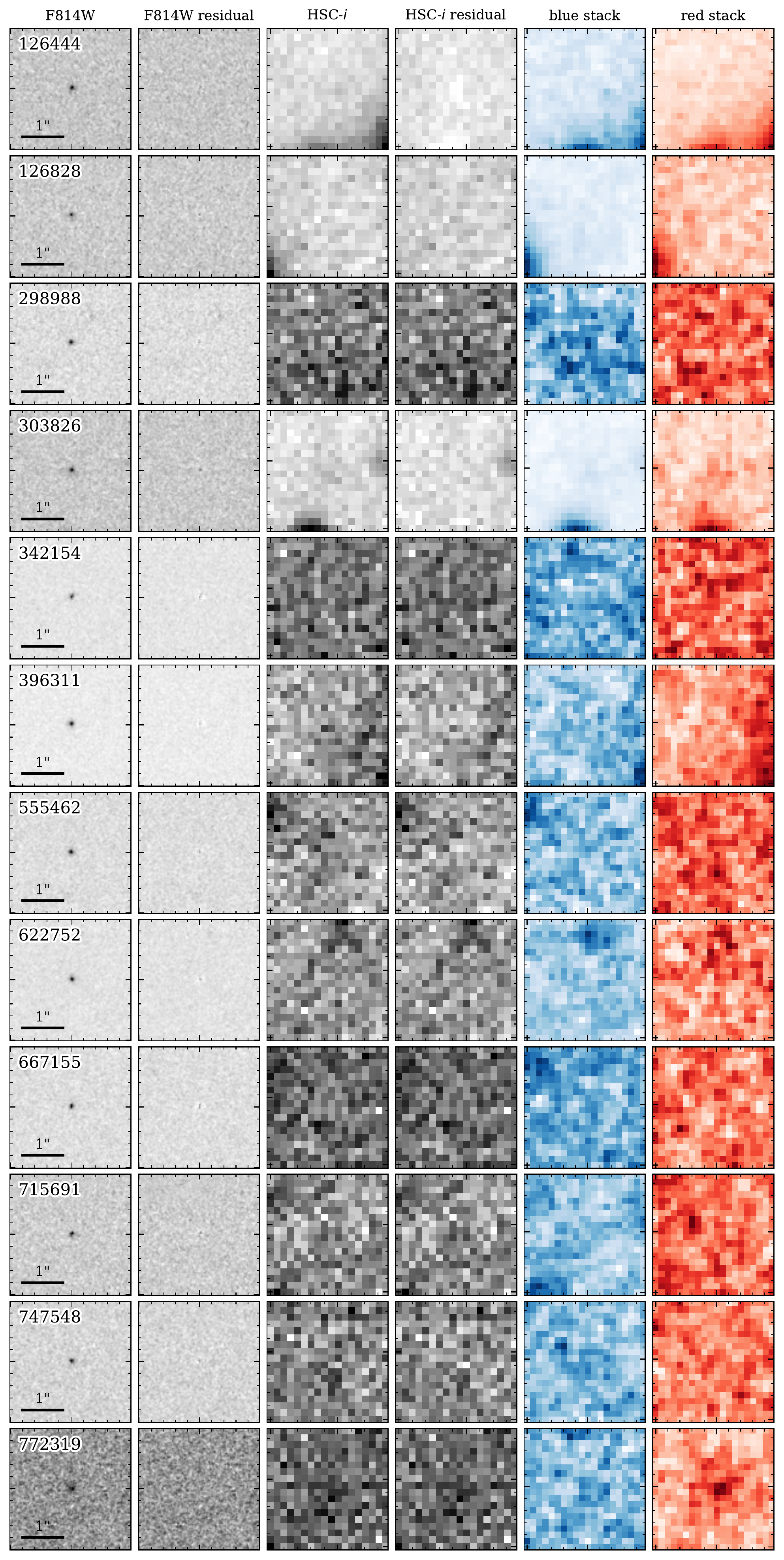}
\end{center}
\caption{Cutouts of the final $12$ candidates (rows). The columns show (from left to write) F814W, F814W residual, HSC-$i$, HSC-$i$ residual, blue stack, and red stack. The cutouts have a size of $3\arcsec\times3\arcsec$, the black bar denotes $1\arcsec$. Only candidate $772319$ is detected in the red stack and has a photometric redshift of $z_{\rm phot} = 5.92$ according to the \textit{COSMOS2020} catalog. None of the candidates are detected in the blue stacks.
\label{fig:cutouts_sample}}
\end{figure*}
%%% ======================

%%% TABLE %%%%%%%%%%%%%
\begin{deluxetable*}{l cccccccc}
\tabletypesize{\scriptsize}
%\rotate
\tablecaption{Summary of candidates.\label{tab:candidates}}
\tablewidth{0pt}
\tablehead{
\colhead{ID} & \colhead{R.A.} & \colhead{Decl.} & \colhead{[F814W]} & \colhead{[HSC-$i$]} & \colhead{[HSC-i]-[F814W]} & \colhead{$R_{\rm e}$} & \colhead{$z_{\rm phot}$} & \colhead{detection in red stack}\\[-0.2cm]
\colhead{} & \colhead{(J2000)} & \colhead{(J2000)} & \colhead{(AB mag)} & \colhead{(AB mag)$^{a}$} & \colhead{(AB mag)$^{b}$} & \colhead{($\arcsec$)$^{c}$} & \colhead{} & \colhead{}
}
\startdata
126444 & 150.6622 & 1.926 & 25.39 $\pm$ 0.02 & 26.58 $\pm$ 0.11 & 1.19 $\pm$ 0.11 & 0.04 & \nodata & no \\
126828 & 150.589 & 1.809 & 25.63 $\pm$ 0.03 & $>$26.90 & $>$1.27 & 0.04 & \nodata & no \\
298988 & 150.4077 & 2.7466 & 25.71 $\pm$ 0.02 & $>$26.90 & $>$1.19 & 0.03 & \nodata & no \\
303826 & 150.3563 & 2.7593 & 25.27 $\pm$ 0.03 & 26.92 $\pm$ 0.17 & $>$1.63 & 0.07 & \nodata & no \\
342154 & 150.1967 & 2.0339 & 25.41 $\pm$ 0.02 & $>$26.90 & $>$1.49 & 0.03 & \nodata & no \\
396311 & 150.2752 & 2.5275 & 25.04 $\pm$ 0.01 & 27.55 $\pm$ 0.26 & $>$1.86 & 0.05 & \nodata & no \\
555462 & 149.8889 & 2.238 & 25.24 $\pm$ 0.02 & 27.59 $\pm$ 0.25 & $>$1.66 & 0.03 & \nodata & no \\
622752 & 149.7843 & 1.7169 & 25.22 $\pm$ 0.01 & $>$26.90 & $>$1.68 & 0.05 & \nodata & no \\
667155 & 149.6293 & 2.2474 & 25.63 $\pm$ 0.02 & $>$26.90 & $>$1.27 & 0.03 & \nodata & no \\
715691 & 149.5934 & 1.6758 & 25.56 $\pm$ 0.02 & $>$26.90 & $>$1.34 & 0.03 & \nodata & no \\
747548 & 149.5442 & 2.2755 & 25.55 $\pm$ 0.02 & 27.73 $\pm$ 0.36 & $>$1.35 & 0.03 & \nodata & no \\
772319 & 149.4493 & 2.5234 & 25.72 $\pm$ 0.01 & $>$26.90 & $>$1.18 & 0.07 & 5.92$^{d}$ & yes \\
\enddata
\tablenotetext{a}{\textsc{Tractor} non-detections have been replaced with the $5\sigma$ limit ($26.9\,{\rm mag}$)}\vspace{-0.2cm}
\tablenotetext{b}{Limits are reported for [HSC-$i$] measurements at $<5\sigma$ and \textsc{Tractor} non-detections}\vspace{-0.2cm}
\tablenotetext{c}{PSF-corrected half-light radii}\vspace{-0.2cm}
\tablenotetext{d}{Has match in the \textit{COSMOS2020} catalog \citep[][]{WEAVER21} }\vspace{-0.2cm}
\end{deluxetable*}
%%%%%%%%%%%%%%%%%%%%%%%%

As an additional check, we compared the sizes of our candidates to the sizes of simulated point sources. For this, we injected $30\,000$ point sources at various magnitudes convolved with the F814W PSF into real ACS F814W images and extracted them using the cataloging procedure.
Figure~\ref{fig:mag_vs_size} shows the results of these simulations on the  \texttt{FLUX\_RADIUS} (half-light radius) versus \texttt{MAG\_AUTO} diagram. Also shown are our $33$ visually inspected candidates. At bright magnitudes, the measurements converge to the half-light radius of the PSF ($\sim2.5\,{\rm pixels}$). At fainter magnitudes, the scatter in half-light radius increases and sizes are generally underestimated due to surface brightness effects, leading to values of $2\,{\rm pixels}$ or less. Our $12$ final candidates (solid blue) are consistent with the sizes of the simulated point sources within their scatter, a further indication that they are real. 

Figure~\ref{fig:cutouts_sample} shows the $3\arcsec \times 3\arcsec$ cutouts in F814W and HSC-$i$ including the corresponding \textsc{Tractor} residuals of the final sample of $12$ candidates. We also show the blue and red stack for each source (see Section~\ref{sec:visualinspection}). None are detected in the blue stack by construction, and candidate $772319$ is the only one detected in the red stack. This candidate (maybe slightly more extended compared to simulated point sources, see Figure~\ref{fig:mag_vs_size}) is also detected in the latest \textit{COSMOS2020} catalog \citep[][]{WEAVER21} and has a reported photometric redshift of $z_{\rm phot}=5.92$. Note that the other candidates are not in the this catalog, which is based only on ground-based imaging. 

Some properties of the $12$ candidates are listed in Table~\ref{tab:candidates}.

%%% FIGURE ===============
\begin{figure}[t]
\includegraphics[angle=0,width=1\columnwidth]{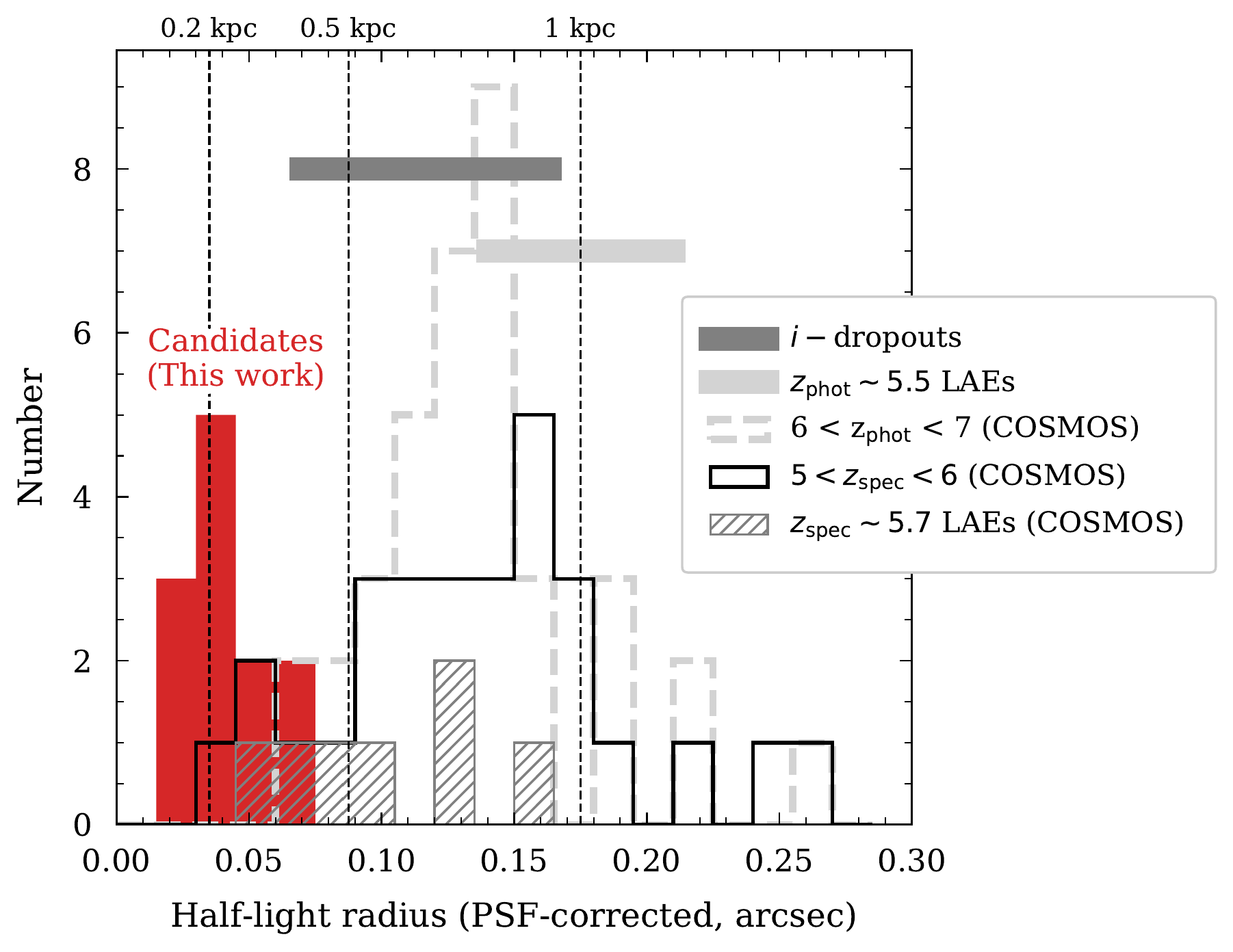}
\caption{Comparison of PSF-corrected sizes of our $12$ candidates (red histogram) with galaxies at similar redshifts selected with various methods (see text for details). We show median sizes and scatter of $i$-band dropouts \citep[dark gray horizontal line][]{MOSLEH12} and narrow-band selected $z\sim5.5$ LAEs \citep[light gray horizontal line][]{PAULINOAFONSO18}, galaxies selected by photometric redshifts at $6 < z_{\rm phot} < 7$ \citep[light-gray dashed histogram][]{WEAVER21}, spectroscopically selected galaxies at $5 < z_{\rm spec} < 6$ \citep[black solid histogram][]{HASINGER18}, and spectroscopically confirmed LAEs at $z_{\rm spec} \sim 5.7$ \citep[light gray hatched histogram][]{SHIBUYA18}. The latter three samples are magnitude matched and in the COSMOS field, hence allows a comparison using a dataset consistent with that used for our candidates. This comparison shows that our candidates are amongst the most compact objects at these redshifts.
\label{fig:size_comparison}}
\end{figure}
%%% ======================

\subsection{Size Comparison to Galaxies at similar redshift}\label{sec:contamin_lbg}

Quasars, outshining their host galaxies, are expected to be unresolved point sources even in observations with space-based observatories such as the HST. In Figure~\ref{fig:size_comparison} we compare the PSF-corrected half-light radii of our candidates (red) to $z\sim6$ galaxies from the literature (gray), which would have similar [HSC-i]-[F814W] colors.
We report the median sizes with scatter of narrow-band selected \lya~emitters (LAEs) at $z\sim 5.5$ \citep[][]{PAULINOAFONSO18} and $i$-band dropouts \citep[$z\sim6$ LBGs][]{MOSLEH12}. Generally, both measurements (especially the one of the LAEs) suggest sizes considerably larger than for our candidates. However, it is to note that while the former study measures the sizes on F814W images, the latter uses WFC3/IR F160W. Furthermore, the images (mostly taken from deep Hubble fields) may differ in depth and reduction from the COSMOS F814W observations. Therefore we also show several samples in the COSMOS field, which allows a direct comparison to our candidates. Specifically, we show the F814W size distributions of photometric galaxies between $6 < z_{\rm phot} < 7$ \citep[][]{WEAVER21}, spectroscopically confirmed $z\sim5.7$ LAEs \citep[][]{SHIBUYA18}, and spectroscopic galaxies between $5 < z_{\rm spec} < 6$ \citep[these include also \lya~undetected galaxies,][]{HASINGER18}. The latter are at slightly lower redshift as we would expect for our candidates. The expected size evolution between these redshift ranges is however less than $10\%$ \citep[see, for example,][]{MOSLEH12}. Note that all these samples are matched in apparent magnitude to our candidates.
This comparison confirms the picture that our candidates, showing sizes of $\sim0.2\,{\rm kpc}$ (assuming $z=6$), are amongst the most compact objects at these redshifts. A population of low-luminosity quasars or exceptionally compact star-forming galaxies (see Section~\ref{sec:discussionSFG}) can explain this.
Note that dust-reddened galaxies at lower redshifts (e.g., $z\sim2$) could show red colors as well. However, such galaxies would be ruled out as they would be resolved in the Hubble images. Furthermore, even extremely dusty low-$z$ galaxies would not be able to produce such a red color in two essentially overlapping bands. For example, to reach a color difference of $0.5\,{\rm mag}$ at $z\sim2$, an obscuration $A_{\rm V}>17\,{\rm mag}$ would be necessary.

%%% FIGURE ===============
\begin{figure}[t]
\includegraphics[angle=0,width=1.0\columnwidth]{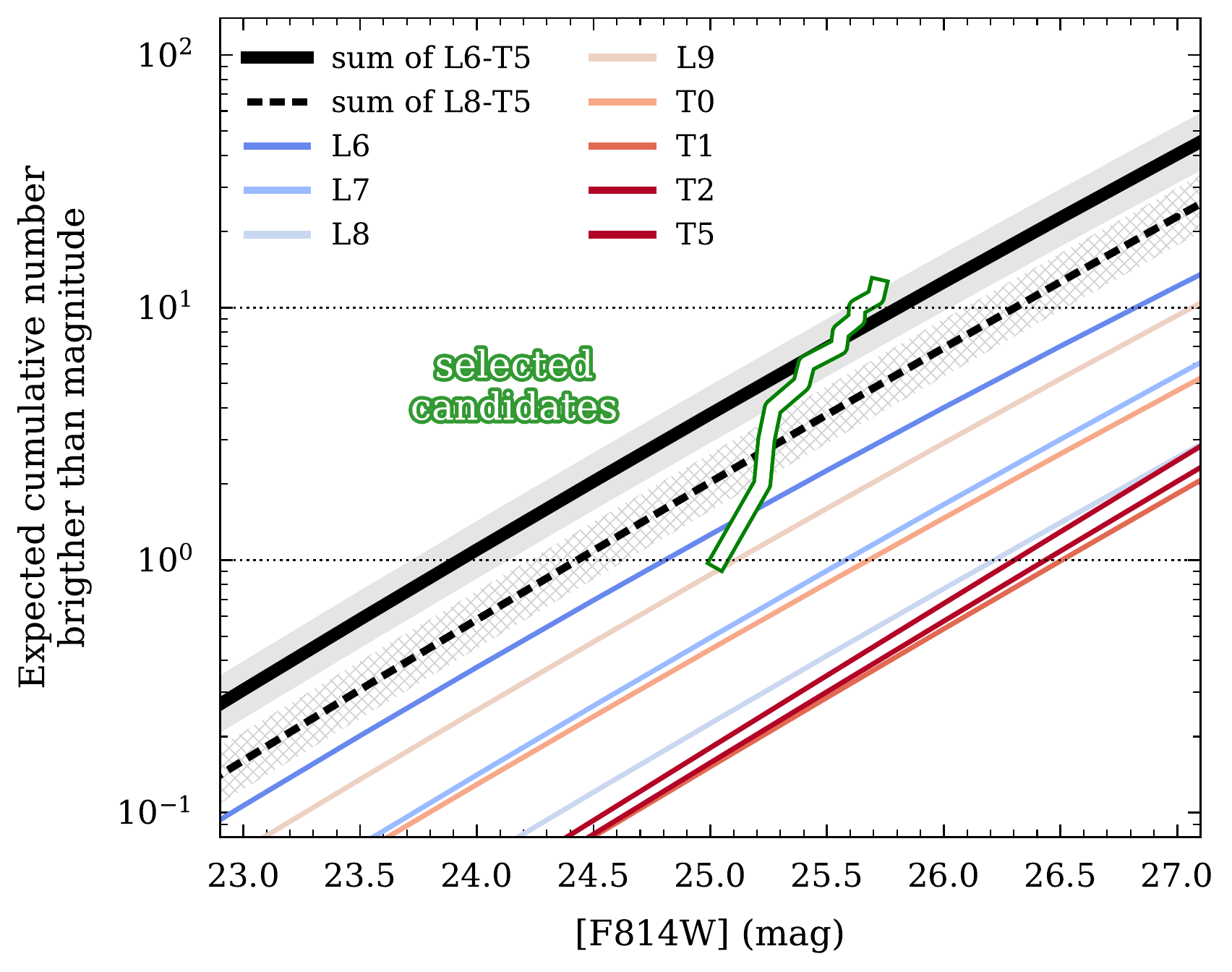}
\caption{Expected cumulative number of brown dwarfs of different spectral types as a function of ACS/F814W magnitude. The total cumulative number of L6-T5 (L8-T6) dwarf stars is shown as a black (dashed black) line together with the uncertainty in number counts from \citet[][]{KIRKPATRICK21}. The horizontal dotted lines show numbers of $1$ and $10$. The green thick line shows the cumulative magnitude distribution of the $12$ final candidates.
\label{fig:LTdensities}}
\end{figure}
%%% ======================

\subsection{Quantifying the Contamination by Stars}\label{sec:contamin_stars}

Late-type stars are likely the major source of contaminants as they can have red [HSC-i]-[F814W] colors and are also unresolved point sources in the HST images. As discussed in the previous sections and shown in Figure~\ref{fig:colorselection}, the applied [HSC-i]-[F814W] color cut of $0.5\,{\rm mag}$ removes stars of spectral type warmer than L5 even for a $V$-band extinction of $2$ magnitudes - we note that this is smaller than the typical Galactic extinction in the COSMOS field. However, the same figure shows that cooler stars can enter our selection.

Removing stars by their proper motion is not possible in our case as our candidates are only detected in one single band and no other deep space-based observations pre or post 2004 (the epoch of the ACS images) are available. At bright magnitudes, Bayesian approaches using variation of light concentration relative to that of the PSF, have been successful at identifying stars but at faint magnitudes, due to surface brightness limitations, these approaches are challenging to apply \citep{SCRANTON2002}.
With the current data in hand, we can therefore only investigate the contamination of stars by quantifying their number densities as a function of spectral type and magnitude.

Figure~\ref{fig:LTdensities} shows the expected cumulative number density of stars with spectral types cooler than L5 over the survey area in the direction of the COSMOS field as a function of F814W magnitude.
The numbers are based on the volume number densities of brown dwarfs in \citet[][]{KIRKPATRICK21} and the absolute magnitudes in the $I$-band from the PanSTARRS ``Three Pi Survey'' \citep[][]{BEST21,BEST17}. We converted the PanSTARRS $I$-band magnitude to ACS/F814W magnitudes for consistency by computing the color of a set of real dwarf stars in the corresponding filters \citep[][]{KIRKPATRICK08,KIRKPATRICK10,KIRKPATRICK11,BURGASSER03,BURGASSER04,BURGASSER06,BURGASSER10}\footnote{See also \url{https://roman.ipac.caltech.edu/sims/Brown_Dwarf_Spectra.html}.}.
For brown dwarfs of spectral types L5 though T5 and apparent magnitudes 18.5 through 28, we computed their surface number densities in a cone towards the COSMOS field subtending $1.64$ square degree and with vertex at Earth. We previously scaled the volume number densities as a function of distance to Earth by using galactic scale height-dependent stellar density profiles of the thin and thick disk components of the galaxy (based on \citet[][]{BUSER00}, \citet[][]{REID93}, and \citet[][]{BINNEY97} Sections 3.6 and 10.4.3.).
The thick green line shows the cumulative F814W magnitude distribution for our final $12$ candidates.

From this figure, we can see that the expected total number of L6-T5 dwarf stars over the surveyed area in our magnitude range is between $5$ and $10$ (thick black line). The largest contribution comes from the warmest stars of L6 spectral type (contributing $1-3$ stars). However, all our candidates have [HSC-i]-[F814W] colors of more than $1$ magnitude (see Figure~\ref{fig:finalselection}), which is too red for L6 and L7 dwarf stars (Figure~\ref{fig:colorselection}). Taking this into account lowers the contamination significantly due to the lower number density of cooler stars at these magnitudes $-$ we expect on the order of $5\pm2$ L8 to T5 stars in our sample (thick black dashed line) for the faintest magnitudes.

\begin{figure}[t]
\includegraphics[angle=0,width=1\columnwidth]{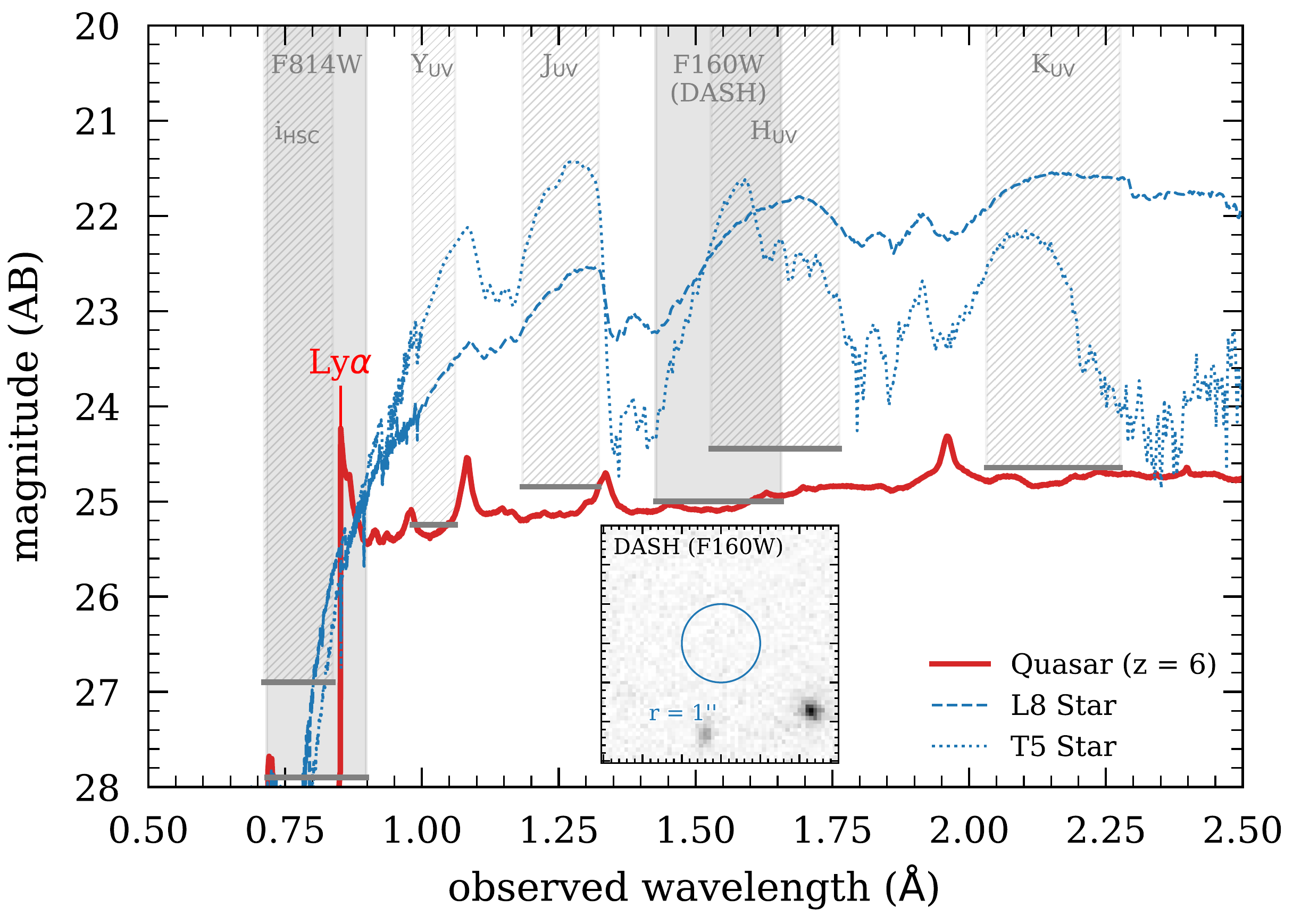}\\
\caption{ If our candidates would be dwarf stars (such as L8 or T5 spectral types, blue dashed and dotted lines, respectively), they should be detected in infrared data such as from the \textit{DASH} HST program ($5\sigma$ point source sensitivity limits indicated by vertical bars). However, for the $7$ candidates with \textit{DASH} observations we do not detect any either individually, or in the stack (inset, blue circle has a radius of $1\arcsec$). The red line shows a quasar template redshifted to $z=6$. All SEDs are normalized to $25.5\,{\rm AB}$ in F814W.
\label{fig:dash}}
\end{figure}

In a next step, we can investigate if such stars are compatible with the non-detections of our candidates in other ancillary data redward of the $i$-band. Figure~\ref{fig:dash} shows a quasar SED \citep[][]{VANDENBERK01} at $z\sim6$ as well as observed L8 and T5 star templates normalized to a magnitude of $25.5\,{\rm AB}$ in the ACS/F814W filter. The vertical bands show the available data and their $5\sigma$ sensitivity limits. Shown are the UltraVISTA $Y$, $H$, $J$, and $K_{\rm s}-$bands as well as the WFC3/F160W ($H$) band from the \textit{DASH} HST program \citep[][]{MOMCHEVA17,MOWLA19}. The latter is available for $7$ of our candidates. The main stellar contaminants would be easily detected in the infrared bands, however, stacking the \textit{DASH} F160W data for the $7$ candidates quasars does not show to a significant detection (see inset). From this, we conclude that stellar contamination is very unlikely for $7$ out of $12$ candidates $-$ for the remaining $5$ candidates we cannot, yet, draw final conclusions.

Summarizing, the red [HSC-i]-[F814W] color cut and the absence of a detection in the \textit{DASH} data suggests that stellar contamination of or sample is very unlikely $-$ at least for $7$ out of our $12$ candidates. However, infrared spectroscopy, for example with \textit{JWST}, will be required for final confirmation.

%%% FIGURE ===============
%%% ======================

\section{Discussion}\label{sec:discussion}

\subsection{Number Density of Low Luminosity $z>6$ Quasar Candidates}\label{sec:discussionLF}

Assuming that all or a fraction of our candidates are true low-luminosity quasars at $z>6$, we can compute their number density and compare it to estimates of the QSO luminosity function derived at higher luminosities.

We derive the number density using a $V_{\rm max}$ method \citep[][]{SCHMIDT68}. Because we do not know the redshifts of our candidates, we carried out a Monte-Carlo sampling of $5000$ galaxies assuming a flat redshift prior between $6 < z < 6.7$ and a Gaussian F814W magnitude distribution (with $\sigma$ equal to the standard deviation of our candidate sample).
Faint galaxies with redshifts close to $z\sim6$ in our Monte Carlo sample ``see'' a smaller volume, hence have higher $1/V_{\rm max}$ (i.e., number densities) associated with them (which causes an asymmetry in the distribution of number densities).

\begin{figure}[t]
\includegraphics[angle=0,width=1.0\columnwidth]{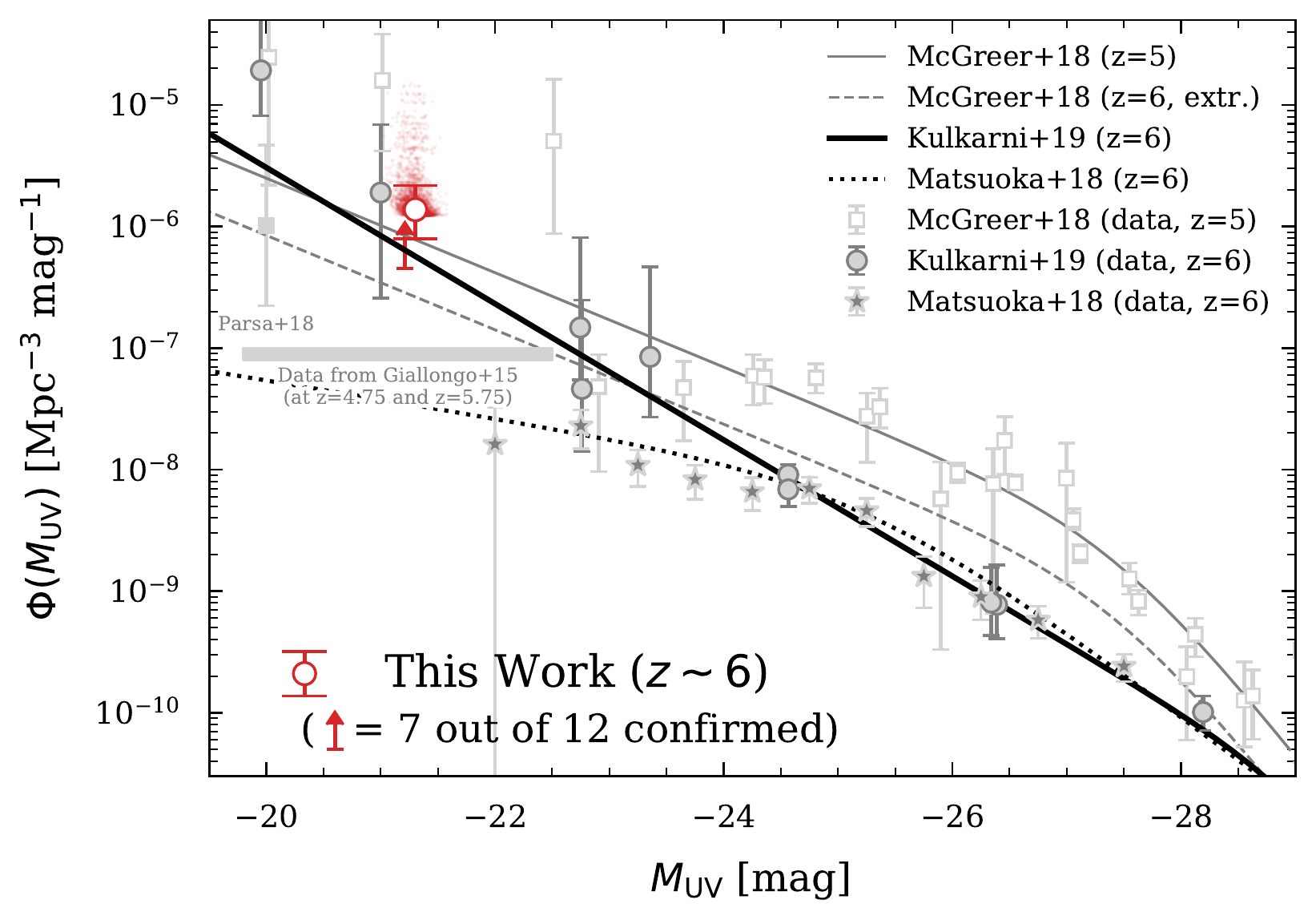}
\caption{Constraints on the faint end of the $z\sim6$ quasar luminosity function by our candidates (red symbols). The distribution from our $V_{\rm max}$ sampling is shown as small red dots. The error bars include the uncertainty from redshift (computed by a Monte-Carlo simulation, see text) as well as shot-noise and cosmic variance ($\sim30\%$). The red arrow shows the lower limit if $7$ out of $12$ candidates are stars (see Section~\ref{sec:contamin_stars}).
The lines show double power-law fits to $z=5$ quasars (gray solid line, data as empty gray symbols) from \citet[][]{MCGREER18}, their extrapolation to $z=6$ gray dashed line) as well as at $z\sim6$ from \citet[][]{KULKARNI19} (black thick line, data as filled gray symbols) and \citet[][]{MATSUOKA18} (spectroscopically confirmed, dotted thick line, data as filled stars). Note that these fits are derived from data brighter than $M_{\rm UV}= -22.8$. The data fainter than that (from \citet[][]{GIALLONGO15} at $z\sim5$ (empty squares) and $z\sim6$ (filled circles), indicated by the horizontal bar) are not used in the fits by McGreer et al. and Kulkarni et al.. For $z=5$, \citet[][]{PARSA18} suggest lower values (filled light gray square).
\label{fig:LFfinal}}
\end{figure}

The resulting distribution is shown as red points together with the median and uncertainty in Figure~\ref{fig:LFfinal}. We find a number density of $1.4^{+0.8}_{-0.6} \times 10^{-6}\,{\rm Mpc^{-3}\,mag^{-1}}$. The error budget includes the uncertainty from the unknown redshift distribution (as derived by our Monte-Carlo simulation), shot noise, as well as the uncertainty introduced by cosmic variance. The latter is estimated by the frame work discussed in \citet[][]{TRENTISTIAVELLI08} using their online cosmic variance calculator\footnote{\url{https://www.ph.unimelb.edu.au/~mtrenti/cvc/CosmicVariance.html}}. We assumed a redshift of $z=6.5\pm1.0$, a square-area of $1.64\,{\rm deg^2}$, a halo filling factor of $1$, and a Press-Schechter bias. We obtain an uncertainty of $30\%$ ($ \pm 5\%$ for different assumptions).
While we have \textit{DASH} HST observations for $7$ candidates, we cannot, yet, rule out stellar contamination for the remaining $5$. Therefore, we also show a lower limit ($0.5\times 10^{-6}\,{\rm Mpc^{-3}\,mag^{-1}}$) in the case of only $7$ confirmed candidates with a red arrow. 

Our estimates are consistent with the extrapolation of the \citet[][]{MCGREER18} luminosity function at $z=5$ (gray solid line) and the \citet[][]{KULKARNI19} luminosity function at $z=6$ (black thick line, data points shown as gray filled symbols) to fainter magnitudes. On the other hand, the $z=5$ \citet[][]{MCGREER18} luminosity function extrapolated to $z=6$ (gray dashed line) underestimates our number density counts by almost a factor of $10$. The study from \citet[][]{MATSUOKA18} at $z\sim6$ shows an even lower number of low-luminosity quasars, which could be due to their spectroscopic selection function. The latter two studies are also not consistent with our lower limit (assuming $7$ confirmed low-luminosity quasars).
Note that these fits shown in Figure~\ref{fig:LFfinal} are heavily constrained by the bright end of the luminosity functions $M_{\rm UV} \leq -22.8\,{\rm mag}$ and thus need to be extrapolated to the luminosities of our candidates.
At fainter magnitudes ($M_{\rm UV} > -22.8\,{\rm mag}$, indicated by the gray horizontal bar), \citet[][]{GIALLONGO15} provide data at $z\sim5$ (empty squares) and $z\sim6$ (filled circles) from putative X-ray detections in the GOODS-S field with photometric redshifts. The measurements of that study at $z\sim5$ are not consistent with McGreer et al. and in fact \citet[][]{PARSA18}, who re-analysed the data from \citet[][]{GIALLONGO15}, suggest that their number densities could be up to a factor of $\sim3$ lower, more in agreement with a turn-over at faint magnitudes of the $z=5$ luminosity function. On the other hand, their number densities at $z\sim6$ are consistent with Kulkarni et al. and our measurement, supporting the steep incline in number density.

%%% FIGURE ===============
%%% ======================

\subsection{Estimated Contribution to Reionization}\label{sec:discussionReionization}

Having placed constraints on the faint end of the QSO LF, we can now calculate how much quasars contribute to the total budget of ionizing photons that led to the reionization of our universe. The answer to this question is still under debate, although it is generally thought that star-forming galaxies are the dominant contributor \citep[e.g.,][]{ROBERTSON15,MADAU15}. The most luminous quasars likely created large bubbles of ionized hydrogen around the most massive dark matter halos \citep[e.g.,][]{MESINGER07,FURLANETTO09}. This likely increased the efficiency with which ionizing photons escape from nearby star-forming galaxies resulting in the further growth of these bubbles to encompass the entire intergalactic medium.
The role of low-luminosity quasars, which reside in lower mass halos, is, however, not clear.
Our measurements add an additional data point to the low-luminosity end of the quasar luminosity function and we can make reasonable assumptions to compute the contribution of such quasars to reionization.

Rather than fitting the data in Figure~\ref{fig:LFfinal}, we adopt the \citet[][]{KULKARNI19} luminosity function parameterization but with a slightly different faint-end slope $\alpha=-2.35$ (instead of $\alpha~=~-2.40$)\footnote{Our limit and scarcity of measurements at $z\sim6$ make the parameterization of the luminosity function rather uncertain. The adjustment of the faint-end slope is therefore done by eye (to match our lower limit) without the use of more sophisticated fitting methods.}, as it matches our limiting case in which only $7$ out of $12$ candidates are confirmed low-luminosity quasars (as we cannot rule out the contribution of dwarf stars for the other $5$, see Section~\ref{sec:contamin_stars}). We find that faint-end slopes of $\alpha = 2.45\pm0.10$ are generally consistent with our measurements.

We integrate the luminosity function described by the (modified) \citep[][]{KULKARNI19} parameters (see above) and find a rest-frame 1500\AA\  luminosity density for these compact sources down to M$_{UV}<-20.5$ mag at $z\sim6.4$ to be $\sim1.1\times$10$^{5}$\,L$_{\sun}$\,Mpc$^{-3}$. We stress that these estimates are lower limits in the case that $7$ out of our $12$ candidates are actual low-luminosity quasars.
Compared to other studies at $z\sim6$, the luminosity densities obtained from our new constraints are similar to those obtained from the \citet[][]{KULKARNI19} and \citet[][]{GIALLONGO15} luminosity function but more than an order of magnitude higher compared to \citet[][]{MATSUOKA18} (see also Section~\ref{sec:discussionLF}).
As reviewed in \citet{Chary_GRBreview}, the star-forming galaxy population produces a rest-frame
UV luminosity density of $\sim$10$^{8}$\,L$_{\sun}$\,Mpc$^{-3}$, which is almost three orders of magnitude larger than the luminosity density estimated in this work. 
If these galaxies indeed show strong nebular line emission as inferred from their multi-wavelength spectral energy distributions, it would imply an ionizing photon production rate which is a factor of 10 higher than among star-forming galaxies in the local Universe. Even with a modest escape fraction of 10\%, such galaxies can easily reionize the IGM by $z\sim6$.
If the Lyman-continuum production rate in the QSO hosts is similar, as seems to be the case based on their inferred nebular emission line properties (B. Lee \& R. Chary, private communication), and the Lyman continuum escape of quasars is not luminosity dependent \citep[as suggested by][]{IWATA22}, it implies that the faint end of the QSO luminosity function has a negligible impact on reionization. 
However, this has the indirect impact of an increased escape fraction of ionizing photons from star-forming galaxies that are within the Str$\ddot{o}$mgren sphere of a given QSO \citep[][]{MESINGER07,FURLANETTO06,FURLANETTO09}).
We therefore conclude that even low-luminosity quasars such as the ones studied here, which are more numerous but several orders of magnitude fainter than the luminous quasars that have been well characterized in earlier studies, play a relatively small role in reionization.
However, low-luminosity quasars can indirectly increase the escape fraction of ionizing photons from star-forming galaxies that are within the Str$\ddot{o}$mgren sphere of a given QSO. This may enhance the role of nearby star-forming galaxies depending on their clustering strength.

\subsection{Alternative Interpretation: Unusually Compact Star Forming Galaxies?}\label{sec:discussionSFG}

Current estimates of the UV luminosity function \citep[e.g.,][]{BOUWENS15}, 
indicate $\sim500$ star-forming $z\sim6$ galaxies in the same area and at the same magnitude as our candidates. As shown in Figure~\ref{fig:size_comparison}, most of these would be resolved in Hubble imaging. However, our candidates, as they are unresolved, could be ultra compact star forming galaxies. In that case, it would be interesting to estimate the star-formation surface density to put them in context with other  
galaxies at $z\sim6$. 

For the estimation of the SFR, the contamination of \lya~to the flux in the F814W filter has to be taken into account. From the F814W magnitudes, we estimate a UV-based SFR for our candidate sample of $70^{+18}_{-10}\,{\rm M_{\odot}\,yr^{-1}}$ using the relation between UV luminosity and SFR in \citet[][]{KENNICUTT98}. This estimate includes the likely contamination by \lya~in the F814W filter, which depends on the EW of the line. We estimate the fraction of flux from \lya~by creating an intrinsic spectrum including \lya~of different EWs, then applying IGM absorption, and finally convolving it with the F814W filter transmission. Assuming an intrinsic EW of $500\,{\rm \AA}$ for \lya~(which is rather a conservative upper limit), we expect \lya~to contribute to about half of the measured UV flux. A more common EW for star-forming galaxies at these redshifts of $<50\,{\rm \AA}$ \citep[see][]{SCHENKER14} results only in $<10\%$ contamination. 

Together with a measurement of a physical size (radius) of our candidates ($\sim0.2\,{\rm kpc}$, see Figure~\ref{fig:size_comparison}), we estimate a SFR surface density of $\Sigma_{\rm SFR}~=~554^{+139}_{-79}\,{\rm M_{\odot}\,yr^{-1}\,kpc^{-2}}$. Including contamination by a \lya~line with EW$=500\,{\rm \AA}$, this number can be up to a factor $2$ smaller. However, as mentioned above, this is a very conservative estimate. Also, note that this estimate describes the unobscured star formation. A moderate dust attenuation of $\ebmv=0.1$ would result in a total SFR density that is a factor $\sim3$ higher. The most likely lower limit for the SFR surface density, is therefore likely around $500\,{\rm M_{\odot}\,yr^{-1}\,kpc^{-2}}$.

We can compare this to the value for a typical main-sequence galaxies at the same redshift. From Figure~\ref{fig:size_comparison}, typical high-z galaxies have a size of around $0.7\,{\rm kpc}$. The average main-sequence SFR (total dust corrected) of a typical $10^9\,{\rm M_{\odot}}$ galaxy\footnote{Since we do not detect most candidates in near-IR bands such as UltraVISTA, we can set an conservative upper limit on stellar mass.} at $z\sim6$ is at $10-100\,{\rm M_{\odot}\,yr^{-1}}$ \citep[e.g.,][]{ALPINE_FAISST20,ALPINE_SCHAERER20,SPEAGLE14}. This results in a SFR surface density of around $5-70\,{\rm M_{\odot}\,yr^{-1}\,kpc^{-2}}$, which is well consistent with the gas mass estimated with ALMA at these redshifts \citep[][]{ALPINE_DESSAUGES20} assuming the Kennicutt-Schmidt relation \citep[][]{KENNICUTT20,KENNICUTT98b}.

This estimate indicates that our candidates, if they are compact star-forming galaxies, show a $7-100 \times$ higher SFR surface densities than typical galaxies at the same redshifts, hence would themselves be interesting targets to follow up spectroscopically.

\section{Conclusions}\label{sec:conclusions}

In this work, we demonstrate the power of a joint pixel-by-pixel analysis of different datasets with different depth and PSF properties. The approach of prior-based photometry mitigates issues of blending and provides robust limits for non-detections. This work is an important stepping stone towards identifying challenges in the analysis steps, and paves the way for applying this method to future datasets such as from the \textit{Rubin}, \textit{Roman}, and \textit{Euclid} missions. A pixel-by-pixel joint analysis of these datasets will result in precision photometric properties and enhance the science output of the individual missions. The \textit{Joint Survey Processing} (JSP) initiative at Caltech/IPAC is working towards building the techniques and infrastructure to carry out such an analysis in the future.

In this work, we have applied JSP to a unique combination of a space and ground-based set of filters in the COSMOS field, resembling in resolution and depth the future imaging data from \textit{Rubin} and \textit{Roman}/\textit{Euclid}. 
We use this technique to
place new constraints on the number density of low-luminosity ($M_{\rm UV} \sim -21\,{\rm mag}$) quasars. 

We perform a pixel-by-pixel analysis of the space-based ACS/F814W filter and ground-based HSC $i$-band filter using \textsc{Tractor}. Specifically,
\begin{itemize}
\item we use the fact that the ACS/F814W filter extends redward of the HSC $i$-band filter, which allows us to select galaxies at $z \gtrsim 6$ using the [HSC-i]-[F814W] colors and;
\item we leverage the high spatial resolution of the space-based F814W filter to select point sources (as expected for quasars).
\end{itemize}

With this technique, we are able to identify $12$ robust candidate sources. Their [HSC-i]-[F814W] colors of $>1\,{\rm mag}$ suggest them to be between $5.9 < z < 6.7$. Their unresolved nature in Hubble imaging (average PSF-corrected half-light radius of $0.03\,\arcsec$ ($\sim 0.2\,{\rm kpc}$ at $z=6$) confirms the compactness expected for quasars.
Number density estimates of L and T dwarf stars identify them as the main likely contaminants (up to $50\%$). However, deep \textit{DASH} HST observations argue against stellar contamination for $7$ of the $12$ candidates.
Taking the potential stellar contamination into account, we estimate a number density of $1.4^{+0.8}_{-0.6}\times 10^{-6}\,{\rm Mpc^{-3}\,mag^{-1}}$ at $M_{\rm UV} \sim -21.2$. If only $7$ out of $12$ candidates are true low-luminosity quasars, we find a $1\sigma$ limiting number density of $0.5\times 10^{-6}\,{\rm Mpc^{-3}\,mag^{-1}}$.
Both numbers may be underestimates. Given the unknown redshift of our candidates, we estimate a potential underestimation of the number densities of a factor $3-4$. Furthermore, the quasar host could dominate the light at these low quasar luminosities, resulting in an extended object \citep[][]{BOWLER21}. Such configurations would be missed as we limit our selection to point sources.

The low number densities of these sources (even including the aforementioned caveats) imply a contribution to reionization which is about three orders of magnitude below that of the star-forming galaxy population. However, low-luminosity quasars may enhance the escape of ionizing photons from neighboring star-forming galaxies and would therefore be interesting environments to study the growth of ionizing bubbles through next-generation spectroscopy.

Alternatively, our candidates could be compact star forming galaxies at $z\sim6$. If true, their SFR surface densities can be estimated to be about $500\,{\rm M_{\odot}\,yr^{-1}\,kpc^{-2}}$, which is $7-100\times$ higher than a typical main-sequence galaxy at these redshifts at $10^9\,{\rm M_{\odot}}$. This would suggest an interesting but rare population of compact star-forming galaxies which future spectroscopy with \textit{JWST} will help characterize.

\software{ \textsc{Astropy 3} \citep[][]{ASTROPY13,ASTROPY18}, \textsc{Tractor} \citep[][]{LANG16a,LANG16b,WEAVER21b}, \textsc{Photutils} \citep[][]{BRADLEY19}, \textsc{SExtractor} \citep[][]{BERTIN96}, \textsc{PSFex} \citep[][]{BERTIN11}, \textsc{TPhot} \citep[][]{MERLIN15,MERLIN16}, \textsc{SkyMaker} \citep[][]{BERTIN09}.}

\acknowledgements
We thank E. Merlin for providing an updated \textsc{TPhot} version and technical support, and the anonymous referee for a constructive feedback on our manuscript. This research used resources of the National Energy Research Scientific Computing Center, a DOE Office of Science User Facility supported by the Office of Science of the U.S. Department of Energy under Contract No. DE-AC02-05CH11231.

The Hyper Suprime-Cam (HSC) collaboration includes the astronomical communities of Japan and Taiwan, and Princeton University. The HSC instrumentation and software were developed by the National Astronomical Observatory of Japan (NAOJ), the Kavli Institute for the Physics and Mathematics of the Universe (Kavli IPMU), the University of Tokyo, the High Energy Accelerator Research Organization (KEK), the Academia Sinica Institute for Astronomy and Astrophysics in Taiwan (ASIAA), and Princeton University. Funding was contributed by the FIRST program from Japanese Cabinet Office, the Ministry of Education, Culture, Sports, Science and Technology (MEXT), the Japan Society for the Promotion of Science (JSPS),Japan Science and Technology Agency (JST), the Toray Science Foundation, NAOJ, Kavli IPMU, KEK, ASIAA, and Princeton University.  

This paper makes use of software developed for the Large Synoptic Survey Telescope. We thank the LSST Project for making their code available as free software at  http://dm.lsst.org

Based in part on data collected at the Subaru Telescope and retrieved from the HSC data archive system, which is operated by Subaru Telescope and Astronomy Data Center at National Astronomical Observatory of Japan.

%% APPENDIX =========
\appendix

%%% FIGURE ===============
\begin{figure}[t]
\centering
\includegraphics[angle=0,width=0.53\columnwidth]{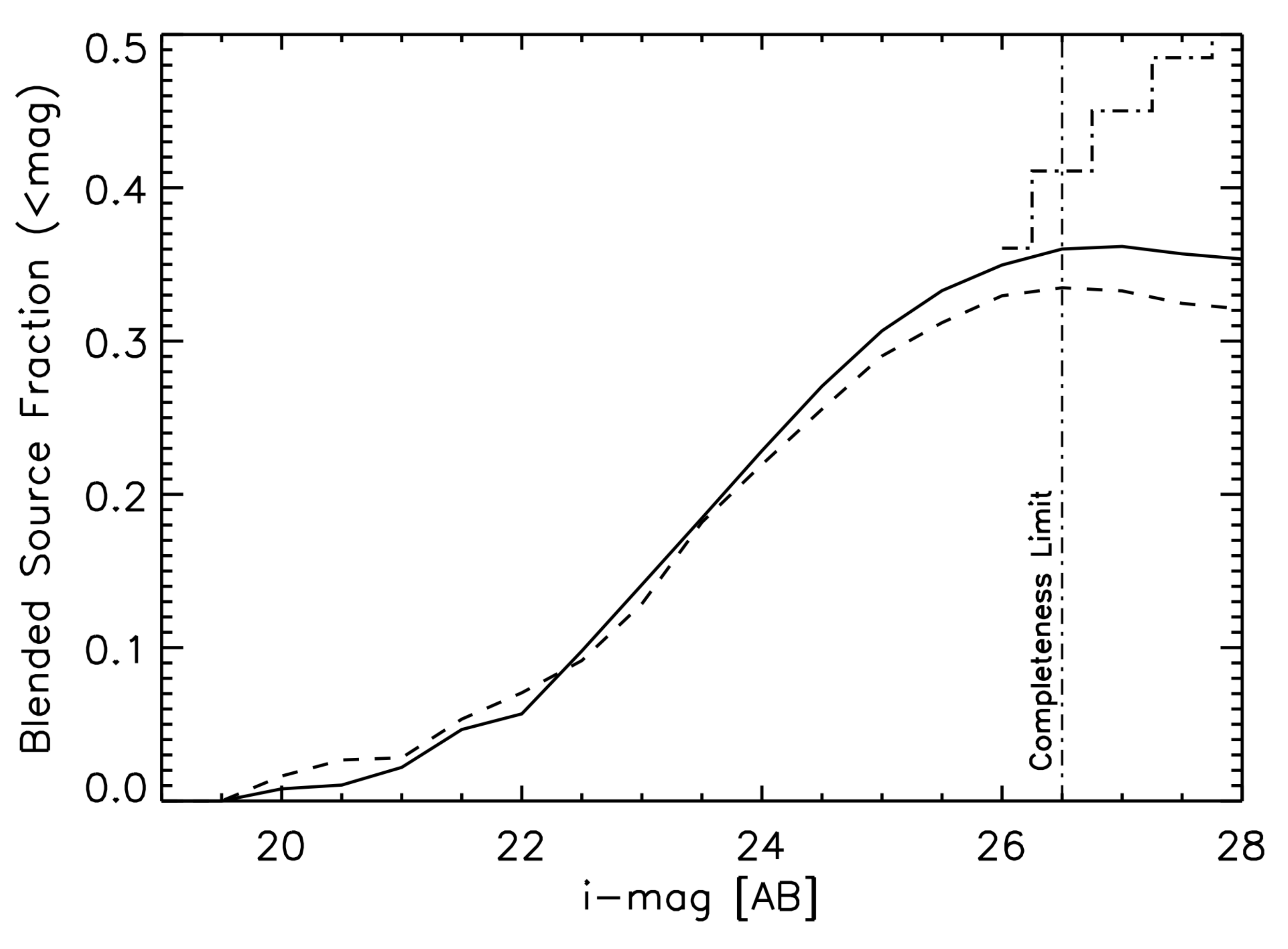}
\includegraphics[angle=0,width=0.46\columnwidth]{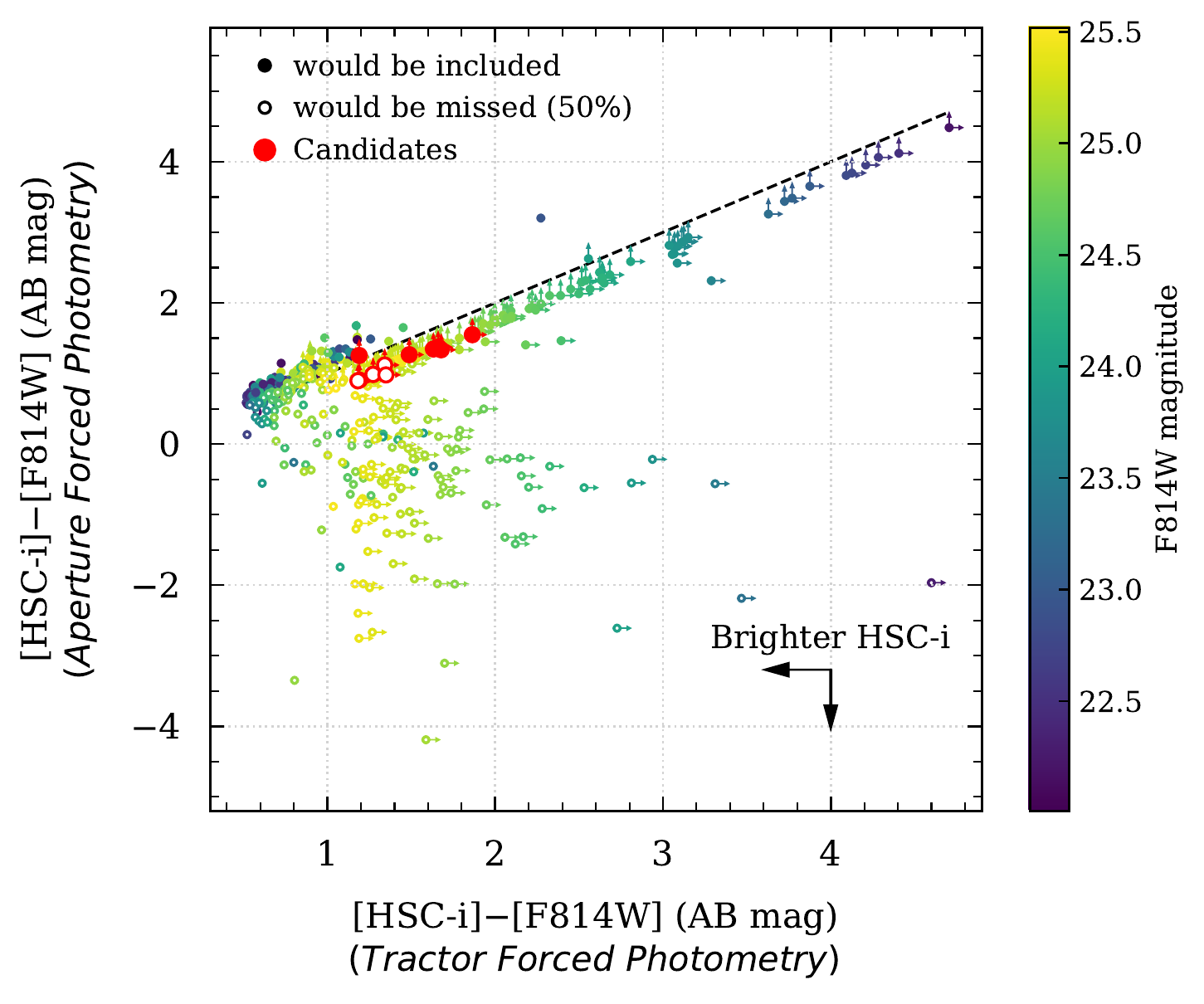}
\caption{\textit{Left}: Fraction of galaxies blended with others as a function of optical $i-$band brightness measured in deep, seeing-limited extragalactic surveys (GOODS-N, solid line, and GOODS-S, dashed line). These catalogs are complete to $26.5\,{\rm mag}$ (vertical line). Numbers beyond that (dot-dashed steps) are found by extrapolation using a power-law fit to the values at brighter magnitudes (see text). At our magnitude limit, we expect $>30\%$ of the sources to be blended. 
\textit{Right:} Comparison of [HSC-i]$-$[F814W] colors of our $555$ initial candidates (see Section~\ref{sec:selection}) measured from \textsc{Tractor} and aperture photometry (see text). The color of each symbol indicates its F814W magnitude.
While the colors of bright objects are consistent between the two methods, the colors (and limits) computed via aperture photometry for fainter objects are significantly bluer in some cases. This is due to flux of neighboring sources bleeding into the aperture around the main target as well as blending of objects within the PSF FWHM. About half of the initial candidates (and $4$ of our final $12$ candidates) would have been missed. This shows the need of prior-based photometry methods (such as \textsc{Tractor} or \textsc{TPhot}) that allow to fit and subtract neighboring sources to provide more robust flux measurements and limits.
\label{fig:colorblending}}
\end{figure}
%%% ======================

\section{Demonstration of Deblending with \textsc{Tractor}}\label{app:blending}

Blending is a significant concern in confusion limited imaging. The left panel of Figure~\ref{fig:colorblending} shows the fraction of blended sources as a function of optical $i-$band brightness in deep, space-based, seeing-limited \textit{CANDELS} fields \citep[GOODS-N and GOODS-S,][]{GROGIN11,KOEKEMOER11}.
Blending was estimated by assuming that the seeing was $0.8\arcsec$ FWHM and if two sources are closer than 1.6", then the photometry of each is biased by more than $5\%$ due to overlapping isophotal areas (assuming Gaussians). The CANDELS catalogs are complete to $26.5\,{\rm mag}$ as shown, so to measure the blending fraction at fainter magnitudes (dot-dashed steps), the source counts were extrapolated using a power law fit to the values at brighter magnitudes. These faint sources were distributed randomly across the sky and the same criterion used to identify the blended fraction. As can be seen in the left panel of Figure~\ref{fig:colorblending}, almost $50\%$ of sources at Rubin/LSST deep survey depths \citep[$i-$band $5\sigma$ magnitude limit over 10 years of $26.8\,{\rm mag}$,][]{IVEZIC19} will be confused and prior-based source fitting has to be undertaken to obtain more robust photometry.

In reality the sources are extended, hence the above estimate is a lower limit.
The general benefit of prior-based photometry over simple catalog matching or aperture photometry is suggested in \citet[][]{WEAVER21} \citep[see also][]{NYLAND17}. Specifically, their study shows an increase in the precision of $\sim20\%$ and slight decrease in outlier fraction of the measurement of photometric redshifts (see their figure~15). However, we note that the \textsc{Tractor} analysis in that study makes use of ground-based priors, which themselves are prone to blending and confusion. The benefit of prior-based photometry is therefore suppressed. A better demonstration is provided in \citet[][]{LEE12}, comparing the HST/F850LP$-$\textit{Spitzer}/$3.6\,{\rm \mu m}$ colors derived from \textsc{SExtractor} (aperture-based) and \textsc{TFit} (the precursor of \textsc{TPhot} using space-based priors) of simulated galaxies. Their figure shows a significant improvement in the measurement of the colors measured with space-based priors.

In the case of our $5\sigma$ HSC $i-$band sensitivity limit, we would expect a $>5\%$ photometry bias in more than $30\%$ of the sources (see left panel of Figure~\ref{fig:colorblending}). 
The use of \textsc{Tractor} (or a similar software such as \textsc{TPhot}, see Appendix~\ref{app:tphot}) allows the mitigation of these effects to provide more robust photometry by using high-resolution prior images and fit neighboring sources simultaneously. Performing prior-based measurement of photometry is especially important in our case, as we are seeking sources that are fainter (or even undetected) on the low-resolution images.

To demonstrate the impact of blending statistically on our results, we compare the [HSC-i]$-$[F148W] colors obtained by our method with simple aperture photometry. The latter would be the choice to search for non-detections $-$ by measuring the flux on the low-resolution image in an aperture around a centroid from a high-resolution image.
In detail, we measure the fluxes in $0.8\arcsec$ diameter apertures on the HSC image at the position of the source detected on the ACS image (thereby we apply the necessary astrometric shifts). We apply an aperture correction, which is calculated by computing the percent of flux of a PSF enclosed within the same diameter. The local background as well as error are calculated from $\sigma-$clipped pixels in an annulus around the source.
The right panel of Figure~\ref{fig:colorblending} shows the results of this comparison for the $555$ initial candidate galaxies that are already color selected (see also Figure~\ref{fig:finalselection}). Generally, sources show aperture-based [HSC-i]$-$[F814W] colors lower then the \textsc{Tractor}-based colors. This is because of flux from neighboring sources ``bleeding'' into the aperture and causing an overestimation of aperture flux (hence bluer color). For isolated and bright sources, both methods return a similar result (within $0.2\,{\rm mag}$).
If aperture-based colors were used, we would miss $50\%$ of potential candidates (open symbols). In addition, four out of our twelve final candidates (red) would have been missed.

%%% FIGURE ===============
\begin{figure}[t]
\centering
\includegraphics[angle=0,width=1\columnwidth]{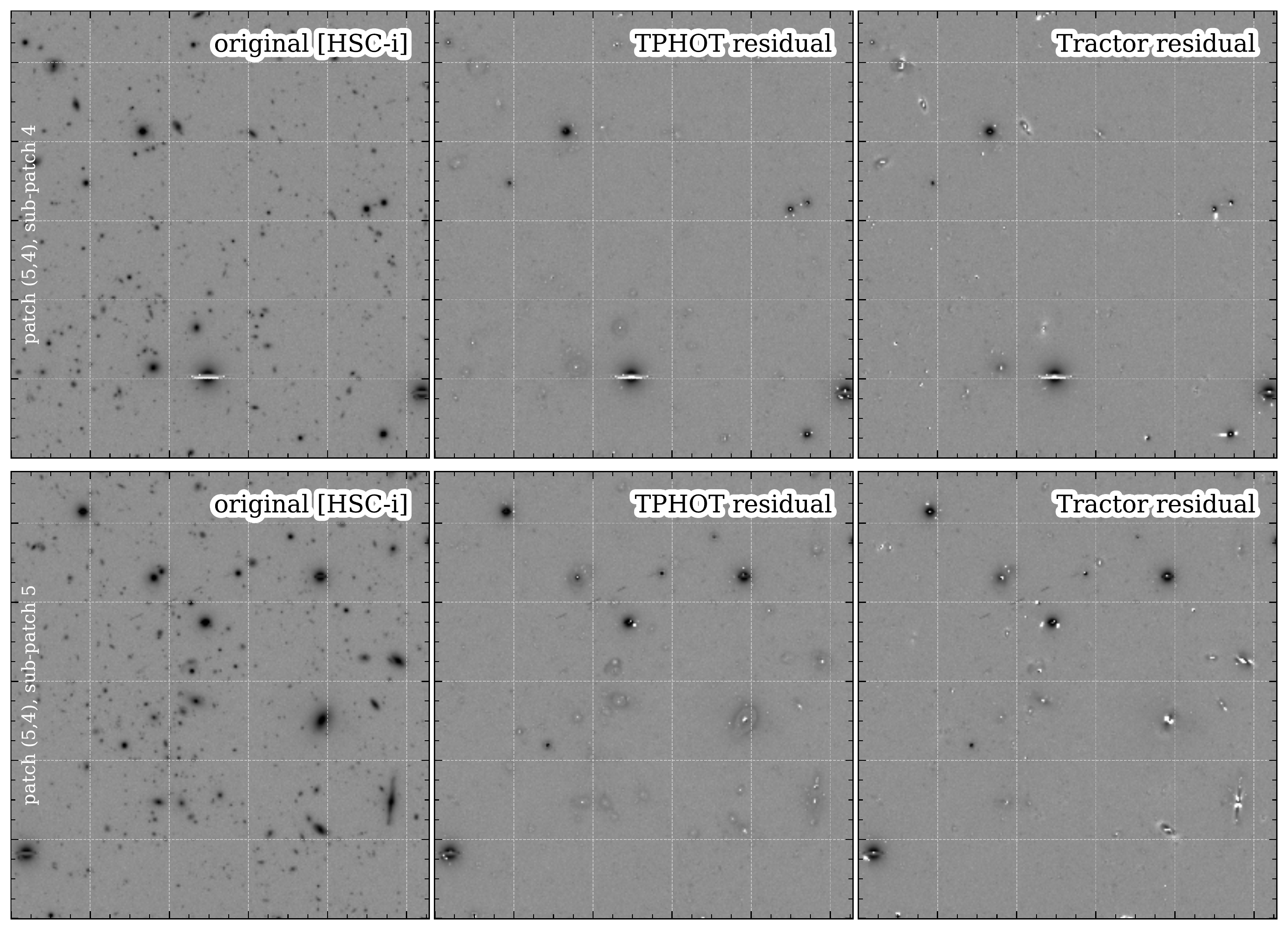}
\caption{Comparison of \textsc{Tractor} (right) and \textsc{Tphot} (middle) residuals of two different sub-patches (top and bottom row).
The original HSC $i$-band image is shown on the left. The scaling is the same in all images.
\label{fig:tphotcomparisonresimage}}
\end{figure}
%%% ======================

%%% FIGURE ===============
\begin{figure}[t]
\centering
\includegraphics[angle=0,width=0.7\columnwidth]{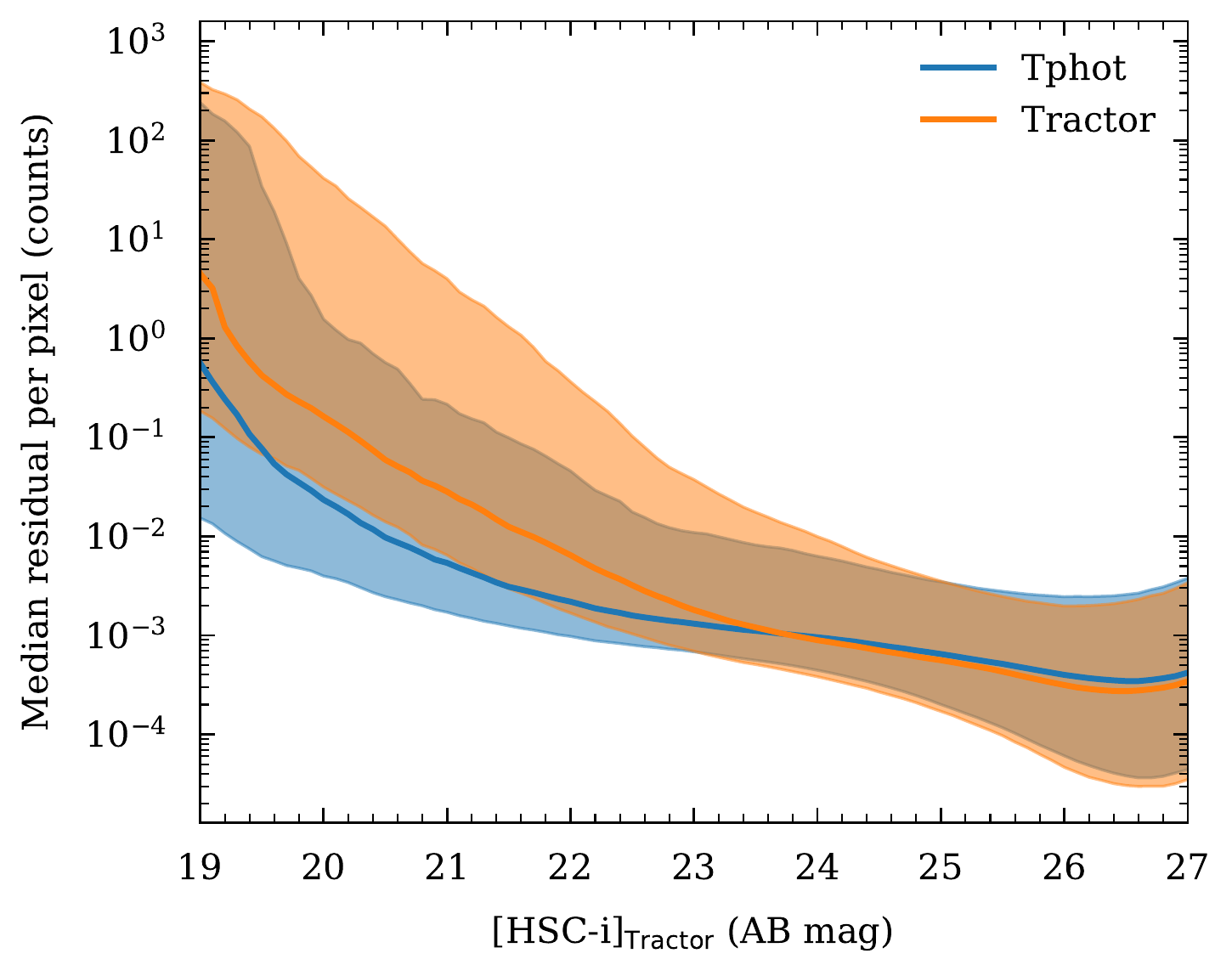}
\caption{Comparison of \textsc{Tractor} (orange) and \textsc{Tphot} (blue) per-pixel residuals as a function of [HSC-i] magnitude. The residuals of both photometry fitting methods are similar at faint magnitudes (preferentially point sources), while \textsc{Tphot} is doing better for bright and extended galaxies due to its non-parametric nature. The shaded area shows the $1\sigma$ range around the median (solid lines).
\label{fig:tphotcomparison}}
\end{figure}
%%% ======================

\section{Comparison to \textsc{Tphot}}\label{app:tphot}

We used \textsc{Tractor} to measure photometry as it allows proper deblending by using high-resolution positional and shape priors. The use of \textsc{Tractor} with simple parametric models (as we do here) has the disadvantage that the models often do not represent the true shapes of galaxies well. This is expected to happen for bright extended galaxies with significant additional sub-structure or lumpy galaxies, for example at high redshifts. In principle, more complicated models can be used at a higher computational cost.
\textsc{Tphot} \citep[][]{MERLIN15,MERLIN16} offers an alternative tool to perform forced photometry measurements. Since it uses non-parametric structural models, we expect it to perform better for bright and extended galaxies.

In brief, \textsc{Tphot} uses the high-resolution image directly by creating cutouts of the galaxies, which are then convolved by a kernel and scaled in flux to fit the galaxies on the low-resolution image. The kernel $K$ (on high-resolution pixel scale) has to be constructed using the low-resolution (lr) and high-resolution (hr) PSF such that ${\rm PSF_{lr} } = K \circledast {\rm  PSF_{hr} }$. This allows \textsc{Tphot} to use the exact shape of galaxies as priors in a non-parametric way.

We ran \textsc{Tphot} on each of the $3\arcmin \times 3\arcmin$ sub-patches. We first identified sources on the high-resolution ACS/F814W image using \textsc{SExtractor}. To account for the different PSF sizes, \textsc{Tphot} creates a dilated segmentation map based on the segmentation map created by \textsc{SExtractor}. We chose the following parameters: \textit{minarea\_dilate}$~=~2.0$, \textit{maxarea\_dilate}$~=~10000.0$, \textit{minfactor\_dilate}$~=~4.0$, \textit{maxfactor\_dilate}$~=~12.0$, \textit{dilation\_factor}$~=~2.0$, \textit{dilation\_threshold}$~=~0.0$.
Because \textsc{Tphot} requires the pixel scale of the low-resolution image ($0.168\arcsec/{\rm px}$) to be an integer multiple of the pixel scale of the high-resolution image, we resized the ACS images from $0.03\arcsec/{\rm px}$ to $0.028\arcsec/{\rm px}$.
To create the convolution kernel, we took the following steps. First, we resized the HSC PSF to the new pixel scale of the high-resolution image using the function \textit{resize\_psf} from the \texttt{PhotUtils} Python package\footnote{\url{https://photutils.readthedocs.io/en/stable/}}. We made sure that the resulting PSF is properly normalized. Second, we used the function \textit{create\_matching\_kernel} (same package) to create the convolution kernel. For the window function, we found that a top-hat filter with an inner diameter of $0.4$ works best. The convolution kernel is verified by convolving the ACS PSF and comparing its profile to the HSC PSF.

Figure~\ref{fig:tphotcomparisonresimage} compares the residuals from \textsc{TPhot} (middle) and \textsc{Tractor} (right) for two different $3\arcmin \time 3\arcmin$ sub-patches. The original HSC $i$-band image is shown on the left. While fainter point sources are fit equally well by both codes, substantial differences in the residuals can be seen for brighter extended galaxies. Specifically, the \textsc{Tphot} residuals are more symmetric while the \textsc{Tractor} residuals show a ``butterfly'' pattern. This is because the former uses the actual shape of the ACS image while the latter is not able to capture the same amount of details in a parametric fit to the galaxies (for example a pronounced bulge $+$ disk or spiral pattern).

Figure~\ref{fig:tphotcomparison} shows this more quantitatively. For each measured source, we computed the per-pixel residual in the same aperture (defined by the segmentation map) on the residual maps produced by the different codes. The figure shows that the residuals left by \textsc{Tractor} are larger at brighter magnitude where more extended galaxies contribute. However, at fainter magnitude (dominated by compact sources), both codes perform equally well.

Summarizing, both codes have advantages and disadvantages. While \textsc{TPhot} shows superior performance by reducing residuals for bright extended galaxies, both codes seem to perform equally well for compact fainter sources ($>23\,{\rm mag})$. However, we found that \textsc{Tractor} is more versatile.
Importantly, \textsc{Tractor} has no constraints on pixel scale or size of the images and no convolution kernel has to be created (which needs some tinkering). 
Furthermore, image artifacts such as diffraction spikes in the vicinity of bright stars affect the functionality of the codes and require dedicated corrections.

\section{Details on the parallelization at NERSC}\label{app:nersc}

An efficient use of supercomputing facilities will be crucial when undertaking joint pixel level analysis of 1000s of square degrees of observations. A single \textit{sub-patch}, $3'\times3'$ in size, with about 1500 sources, requires about 20--50 minutes to run on a high-end 2019 laptop or desktop computer.  Scaling this up to something as small as the COSMOS field would require about two weeks to one month of computing time on a single desktop machine for a single pair of wavelengths. Joint pixel analysis is however intrinsically parallelizable.
We made use of the supercomputer \textit{Cori} at the National Energy Research Scientific Computing Center (NERSC\footnote{\url{https://www.nersc.gov/}}) to execute our pipeline.

The pipeline, written in Python, was containerized using \textsc{Docker}\footnote{\url{https://www.docker.com/}} and then deployed on \textit{Cori} using \textsc{Shifter}\footnote{\url{https://www.nersc.gov/research-and-development/user-defined-images/}}, which is a custom software containerization solution, developed by NERSC. \textsc{Shifter} transforms standard \textsc{Docker} images into a custom format which can be used to automatically launch containers on the nodes of the supercomputer.

To deploy the pipeline on multiple nodes, we made use of a \textsc{Slurm}\footnote{\url{https://slurm.schedmd.com/}} batch script, as shown below.

\begin{lstlisting}[language=Bash]
#!/bin/bash
#SBATCH --job-name=JSP
#SBATCH --account=xxx
#SBATCH --time=48:00:00
#SBATCH --constraint=haswell
#SBATCH --qos=regular
#SBATCH --mail-type=BEGIN,END,FAIL,TIME_LIMIT,TIME_LIMIT_80
#SBATCH --image=docker:nrstickley/jsp_apps:2019-12-20
#SBATCH --nodes 1
#SBATCH --exclusive
#SBATCH --array=0-27
#SBATCH --output=%x_%A_%a.out

shifter /bin/bash run_jobs.sh ${SLURM_ARRAY_TASK_ID} ${SLURM_CPUS_ON_NODE}
\end{lstlisting}

This allocates an array of 28 computing nodes and launches the script \texttt{run\_jobs.sh} in a \textsc{Shifter} container on each. Each node is labeled with integer ID number from 0 to 27, which is stored in the \texttt{SLURM\_ARRAY\_TASK\_ID} variable. The number of logical CPU cores per node (64, in the case of Cori) is passed to the script using the variable, \texttt{SLURM\_CPUS\_ON\_NODE}. The total time allocation per node was set to the maximum of 48 hours.

The most compute-intensive component of the pipeline, \textsc{Tractor}, is multi-threaded but scales poorly with core count. Running multiple instances of \textsc{Tractor} on each node in parallel improves efficiency, but memory consumption becomes the limiting factor. The maximum memory needed by a single task was slightly less than 16~GB. Since each Cori compute node contained 128~GB of system memory, we were able to run 8 simultaneous tasks per node. This was done using the script \texttt{run\_jobs.sh}, which used \textsc{GNU Parallel} \cite[][]{TANGE11a} to efficiently distribute multiple simultaneous instances of our code across the CPU cores of each individual compute node.

\begin{lstlisting}[language=Bash]
#! /bin/bash
# run_jobs.sh runs multiple jobs on a single node.

NODE_ID=${1}  # node ID, which is the SLURM_ARRAY_TASK_ID
SLURM_CPUS_ON_NODE=${2}  # the number of logical CPUs on the node.

# define number of simultaneous jobs
njobs=8

cpus_per_job=$((SLURM_CPUS_ON_NODE / n_jobs))

echo Task ID is: ${SLURM_ARRAY_TASK_ID}
 
ls /path/${NODE_ID}/*.json | parallel -j ${njobs} ./task_runner.sh ${cpus_per_job} {%} {}

\end{lstlisting}

The \texttt{run\_jobs.sh} script operates on a list of JSON-formatted job definition files, each of which contains all of the input parameters that our pipeline (Section~\ref{sec:pipeline}) needs to run a specific \textit{sub-patch}. There are 1008 job files, which we have distributed into 28 directories (named 0 to 27 corresponding to the node ID, \texttt{SLURM\_ARRAY\_TASK\_ID}). \textsc{GNU Parallel} dynamically schedules the tasks into 8 execution `slots' so that as soon as one task completes, another one is launched in the same slot. This continues until the full list of tasks assigned to the node has been executed.

The \texttt{run\_jobs.sh} script, launches an instance of \texttt{task\_runner.sh} for each job assigned to the node. The script, \texttt{task\_runner.sh} handles the details of running an individual task on a set of CPU cores. 

\begin{lstlisting}[language=Bash]
#! /bin/bash
# task_runner.sh distributes the work over 8 physical CPU cores on a single NUMA node.

cores_per_task=$1  # the number of logical cores per task
slot=$2 # The execution slot; an integer from 1 to 8
filename=$3 # JSON job file for one 3' x 3' sub-patch

# This function determines which CPU threads correspond to the specified execution slot:
cpu_threads() {
    minus=$((slot - 1))
    first_cpu_thread=$((cores_per_task * minus))
    final_cpu_thread=$((first_cpu_thread + cores_per_task - 1))
    threads=$(seq $first_cpu_thread $final_cpu_thread | awk -vORS=, '{print $1}')
    echo ${threads::-1}
}

taskset -c $(cpu_threads) python runTractor.py ${filename}
\end{lstlisting}

The utility \texttt{taskset} specifies which specific logical CPU cores will be used to execute the \texttt{runTractor.py} script. This forces all of the threads of the task to remain on the same set CPU cores for the duration of their lifetime, rather than migrating to other cores over time, which improves the cache utilization and reduces the memory access latency. For the first execution slot, these would be logical cores 0-7. For the second slot, 8-15, and so on.

In summary, by distributing the jobs over 28 compute nodes, with each node running 8 jobs simultaneously, we were able to complete all 1008 tasks in slightly less than 24 hours. We plan to implement additional optimizations in the future. In particular, it is possible to run more tasks simultaneously on a single compute node by more intelligently scheduling the tasks to avoid running out of memory.

%% For this sample we use BibTeX plus aasjournals.bst to generate the
%% the bibliography. The sample63.bib file was populated from ADS. To
%% get the citations to show in the compiled file do the following:
%%
%% pdflatex sample63.tex
%% bibtext sample63
%% pdflatex sample63.tex
%% pdflatex sample63.tex
\bibliographystyle{aasjournal}
\bibliography{bibli}

%% This command is needed to show the entire author+affiliation list when
%% the collaboration and author truncation commands are used.  It has to
%% go at the end of the manuscript.
%\allauthors

%% Include this line if you are using the \added, \replaced, \deleted
%% commands to see a summary list of all changes at the end of the article.
%\listofchanges

\end{document}